\documentclass[12pt,preprint]{aastex}

\newcommand{\ds}{$\delta$ Scuti{ }}

\newcommand{\Sec}{${}^{\prime\prime}$\llap{.}}
\newcommand{\Min}{${}^{\prime}$\llap{.}}
\renewcommand\vr{\mbox{$(V\!-\!R)${ }}}

\newcommand{\udeg}{${}^{\textrm{o}}$\llap{.}}

\shorttitle{Time-Series Photometry of M11}
\shortauthors{Hargis et al.}

\begin{document}
 
\title{Time-Series Ensemble Photometry and the Search for Variable
Stars in the Open Cluster M11}

\author{Jonathan R. Hargis\altaffilmark{1}} 
\affil{San Diego State University, Department of Astronomy, 5500 Campanile Drive, San Diego, CA 92182} 
\email{jhargis@eastern.edu}

\author{Eric L. Sandquist}
\affil{San Diego State University, Department of Astronomy, 5500 Campanile Drive, San Diego, CA 92182}
\email{erics@sciences.sdsu.edu}

\author{David H. Bradstreet}
\affil{Eastern University, Department of Physical Sciences, 1300 Eagle Road,
  St. Davids, PA 19087-3696}
\email{dbradstr@eastern.edu}

\altaffiltext{1}{Present address: Eastern University, Department of
  Physical Sciences, 1300 Eagle Road, St. Davids, PA 19087-3696}

\begin{abstract}
This work presents the first large-scale photometric variability
survey of the intermediate age ($\sim200$ Myr) open cluster
M11. Thirteen nights of data over two observing seasons were analyzed
(using crowded field and ensemble photometry techniques) to obtain
high relative precision photometry.  In this study we focus on the
detection of candidate member variable stars for follow-up studies.  A
total of 39 variable stars were detected and can be categorized as
follows: 1 irregular (probably pulsating) variable, 6 \ds variables,
14 detached eclipsing binary systems, 17 W UMa variables, and 1
unidentified/candidate variable.  While previous proper motion studies
allow for cluster membership determination for the brightest stars, we
find that membership determination is significantly hampered below
$V=15,R=15.5$ by the large population of field stars overlapping the
cluster MS.  Of the brightest detected variables that have a high
likelihood of cluster membership, we find five systems where further
work could help constrain theoretical stellar models, including one
potential W UMa member of this young cluster.
\end{abstract}

\keywords{binaries: eclipsing --- delta Scuti --- open clusters and
associations: individual (M11)}

\section{Introduction\label{intro}}

Our Galaxy's collection of open clusters presents an important
population of stars for many areas of modern astrophysics, in
particular for studies of variable stars and extrasolar planets.
Because of the characteristics common to stars in open clusters
(namely the age and chemical composition), much more information can
be obtained for a particularly interesting cluster member star than
could be otherwise deduced for an isolated field star.  The high
spatial density of stars in clusters allows for the opportunity to
survey many stars at one time, a property important to increasing the
probability of detection for extrasolar planetary transits.  Other
advantages of open cluster surveys for extrasolar planets have been
summarized by \citet{vb04}.  Observations of the various types of
variable stars in clusters can provide the other critical remaining
parameters necessary for studying the physics of stars: masses, radii,
and luminosities.  In the case of eclipsing binary systems and
extrasolar planets, photometric observations of the eclipses yield
constraints on the the orbital inclination and relative radii, the
necessary compliments to radial velocity measurements that allow for
the determination of the absolute system parameters.  With the
absolute properties measured, strong tests of structure and evolution
become possible for both extrasolar planets \citep{ba03,ch04,bu04} and
stars (for example Ribas 2003, Lacy et al. 2004).

In this paper we focus on the search for variable stars in the open
cluster M11, identifying systems important for follow-up studies.  In
$\S\ref{target}$ we introduce the target and previous studies of
this object.  In $\S\ref{obser}$ we describe the observations of the
cluster.  $\S\ref{analysis}$ details the data analysis, photometry,
and search for variable stars.  In $\S\ref{CMD}$ we describe the
cluster color-magnitude diagram (CMD).  The light curves for the
detected variable stars are presented in $\S\ref{lcs}$.  We give the
conclusions and future work in $\S\ref{conc}$.

\section{The Target Cluster: M11\label{target}}

The open cluster M11 (NGC 6705, C 1848-063) is a rich and dense open cluster,
containing on the order of several thousand stars.  It is both
relatively young and metal rich (see below), both of which are
advantageous for searches for planetary transits; younger planets will
be larger on average (causing a deeper transit, being more easily
detectable) and have been observed to show a preference for metal-rich
host stars \citep{sa03}.  Here we present an overview of some
relevant previous studies of M11 and the known variable stars in M11.

\subsection{Previous Studies of M11}

M11 has historically been the focus of cluster dynamics studies, but
little work has been done on the variable star content.  The earliest
significant photometric study of M11 was conducted by \citet{jo56},
who derived a CMD from the $UBV$ photographic data for stars brighter
than $V$=15.  They were the first to note the detection of a
population of approximately 30 red giant stars, and found the cluster
age to be intermediate between Praesepe and the Pleiades.  Recent
estimates of the age of M11 by \citet{su99} find an age of
$(200-250)\times10^6$ yr.  A study of the cluster dynamics for stars
brighter than $V\sim15$ magnitude was done by \citet{ma77} based on
the proper motion study by \citet{mps77}.  These studies also noted
the presence of a halo of low-mass stars, which was later confirmed by
\citet{so80}.  \citet{ma84} performed further photometry of M11 to
form a complete and consistent sample of photometric measurements to
compliment the proper motion measurements of \citet{mps77}.  This
study also examined, in detail, the cluster structure and dynamics of
M11.  It was not until the study by \citet{br93} that the cluster CMD
was observed below $V\sim15$; observations reached a limiting
magnitude of $V\sim22$.  The most recent photometry has been presented
by \citet{su99}, who covered an area of approximately
40\Min0$\times$40\Min0 around the center of M11.  They derive an
unreddend distance modulus of $(V_o-M_v)=11.55\pm0.10$ (distance =
2 kpc) and cluster radius of 16 arcminutes (= 9.5 pc for a distance
of 2 kpc).  Spectroscopic measurements have also been made of the red
giant branch population of M11.  \citet{go00} presented an abundance
analysis of 10 red giant stars in M11, and deriving a metallicity of
[Fe/H]=$+0.10\pm0.14$ from high-resolution spectra.

In the galaxy, the cluster is located ($l=$27\udeg3,$b=-$2\udeg8) near
the Scutum star cloud and the Sagittarius-Carina arm, resulting in a
very large contamination of the cluster CMD from the field population
\citep{su99}.  As first noted by \citet{ma84}, the CMD of the M11
field appears to have not only the typical red field star population,
which is easily distinguished from the cluster main-sequence (MS), but
also has a population of stars overlapping the cluster MS starting at
approximately $V=15$ and fainter.  This has also been noted by
\citet{br93}, but is most clearly seen in Fig. 10 of \citet{su99}.
Here, the population of stars is shown divided into 4 different
sections: (1) the MS of M11, (2) the ``blue'' field star population,
(3) field (lower division) and cluster (small upper division) giant
stars, and (4) the ``red'' field star population.  It is the
population of field stars from group \#2 that hampers membership
estimation for many of the detected variables in this study.  Without
direct distance measurements, it is difficult to say with certainty
that a given star falling in the overlap region (sections \#1 and \#2)
is either a cluster member or member of the bluer field star
population.  As further confirmation that the stars in section \#2 are
not cluster members, Sung et al. (1999, Figure 5) derive a CMD for a
region nearby M11, which clearly shows the bimodal field star
population.

\subsection{Known Variable Stars in M11\label{known}}

Observations of photometrically variable stars in this cluster and its
vicinity have been rare.  To date, no comprehensive study of the
variable star population has been published. The most well-studied
potential cluster member is BS Scuti, an Algol-type eclipsing binary
($P=3.8$ d; \citealt{ha74}). The star IT Scuti, located much closer to
the cluster center, is characterized as a slow irregular variable,
although no light curve has been published.  The only previously
known, confirmed \ds variable is V369 Scuti (\citealt{mps77} ID \#624;
$P=0.223$ d; \citealt{ha70}).  A recent survey for variability in M11
was performed by \citet{pa04}, but the search concentrated on the
brightest stars only.  They note the detection of one variable star
(ID \#770 from \citealt{mps77}) but do not indicate the type of
variability detected nor present a light curve.  Lastly, two
spectroscopic binaries (IDs \#926 and \#1223, discovered in a radial
velocity survey by \citealt{ma86}) were analyzed by \citet{le89}.  We
discuss our observations of these previously known and any suspected
variable stars in $\S\ref{detection}$.

\section{Observations\label{obser}}

The data for this study were obtained at the 1 m telescope at the
Mount Laguna Observatory using a $2048\times2048$ CCD.  In total, M11
was observed over the two observing seasons of 2002 and 2003 and 13
nights of data employed in this study are presented here.  Only
observations where the nightly transparency (weather conditions) were
good/excellent or photometric were used in this study.
Table~\ref{obs} presents the log of the observations; observing
conditions are also noted, with error estimates to give some
indication as to the photometric stability of the nightly data.  Given
a CCD plate scale of 0\Sec4 pixel$^{-1}$, the total surveyed area
around the cluster center was 13\Min7$\times$13\Min7.  In radial
extent we have observed out to a radius of 6\Min8 from the cluster
center; our observations have primarily covered the central portion of
the cluster.  Time-series observations (exposure times of 500 s were
used in all monitoring data) were done exclusively in the $R$ band to
maximize the number of surveyed stars.  Owing to the large readout
time of the CCD, there was a delay of approximately 6.7 minutes
between exposures.  Observations were made on 02-03 September 2002 in
the $V$ band (exposure times of 10 s, 60 s and 300 s) for use in the
construction of the \vr color index.

\section{Data Analysis\label{analysis}}

\subsection{Data Reduction}

Observations of the target cluster were reduced in the standard
fashion, employing the IRAF\footnote{Image Reduction and Analysis
Facility is distributed by the National Optical Astronomy
Observatories.} routines to correct the raw data.  The bias level was
subtracted using a fit to the overscan region of the CCD frames, and
the overall noise level was further adjusted using a set of master
bias frames.  Pixel-to-pixel sensitivity variations were corrected
through the use of flat fielding.  Twilight flats were used where
possible, and otherwise dome flats were employed.

\subsection{Photometry}

Photometry was performed using the DAOPHOT II/ALLSTAR suite of
programs \citep{pbs87}. Typically 80-100 bright, well-isolated stars
were selected in construction of the frame point-spread function
(PSF).  The selection was also constrained by the division of the
frame into 25 spatial bins, ensuring that the PSF stars were
adequately spread across the frame (aiding in mapping any possible PSF
variations across the image).  In order to perform consistent
photometry from night-to-night, one master star list of 11,267 stars
(constructed from the best frames of 08/09 August 2002) was used
uniformly throughout this study and only these stars were studied for
variability.  The large numbers of stars on each frame ensured
excellent frame-to-frame positional transformations, with typical
transformations being accurate to better than 0.1 pixels.
Night-to-night differences in the frame center and seeing conditions
hinders measurements of every master star in every frame, but
typically 9,000-10,500 stars were measured in each night of data.

\begin{figure}
\plotone{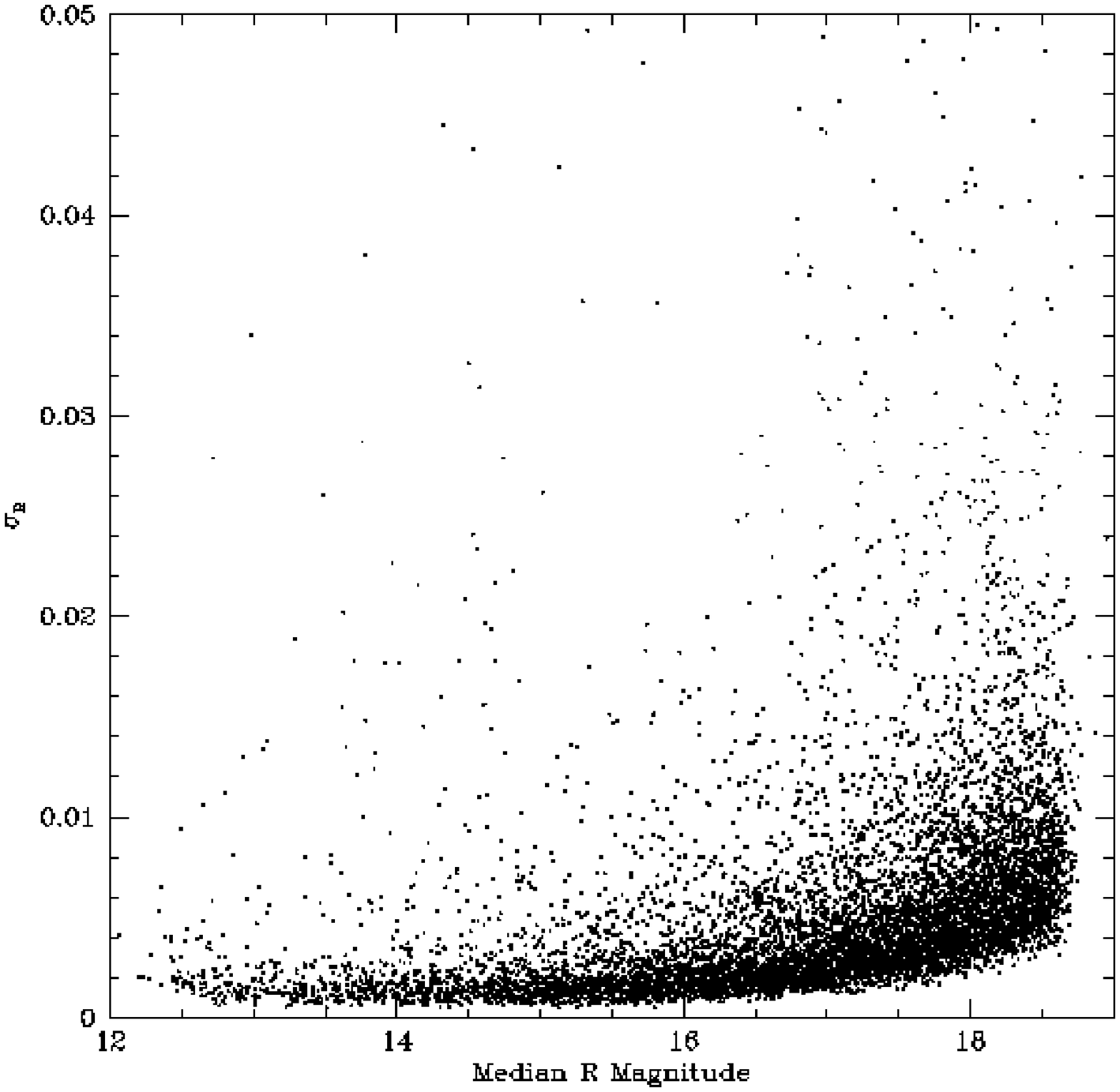}
\caption{Results of the ensemble procedure used to interatively
determine the nightly zero-points and median magnitudes.  Shown is the
error in the median $R$ magnitude $\sigma_{R}$ as a function of median
$R$ magnitude.\label{errors}}
\end{figure}

In order to improve the relative photometry for the light curves,
ensemble techniques \citep[based on a general method by
\citealt{ho92}]{sand03} were used to determine a simultaneous, robust
solution for the median magnitudes of all stars and the relative frame
zero points.  The initial solution resulted from the star-to-star
comparisons determined from the positional transformations, and the
solution was improved via iteration until neither the frame zeropoints
nor the median magnitudes varied by more than 0.0003 magnitudes.  Also
included in the iterative procedure was the possibility that the
magnitude residuals could be a function of frame $x$ and $y$ position.
Second-order polynomials were fit to the magnitude residuals and the
solution was subtracted during the iterations.  At most, these
corrections were only a few hundredths of a magnitude.
Figure~\ref{errors} displays the results of the ensemble techniques,
showing the error in the median $R$ magnitude as a function of median
$R$ magnitude.

\subsection{Variable Star Detection\label{detection}}

\begin{figure}
\plotone{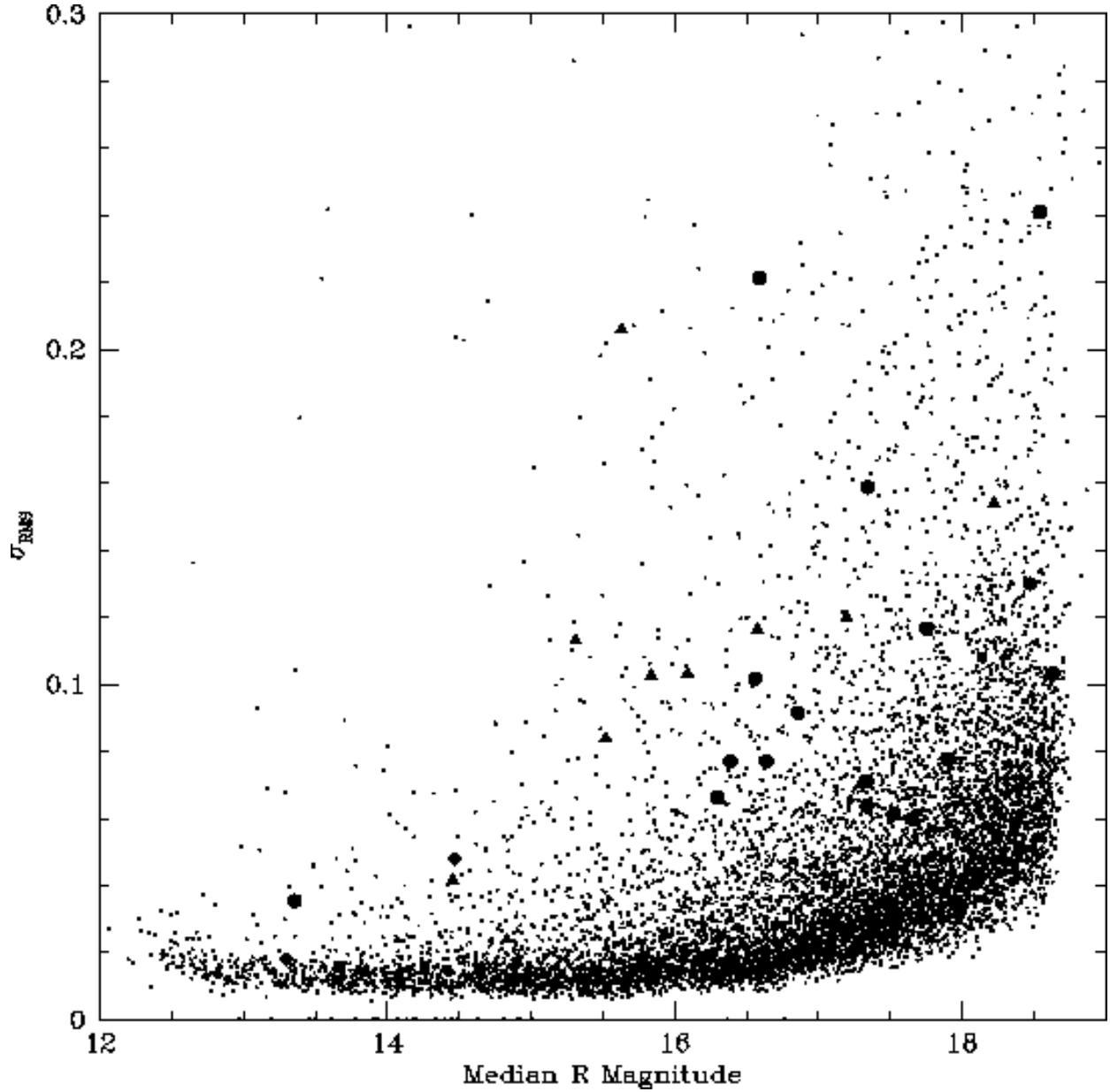}
\caption{Results of the calculation of the RMS variation $\sigma_{RMS}$
about the median $R$ magnitude for the total data set.The detected
variables are denoted as follows: \textit{Filled Square}-\ds or
pulsating variable, \textit{Filled Circle}-W UMa variable,
\textit{Filled Triangle}-detached eclipsing binary, and \textit{Filled
Diamond}-unknown/unidentified variable.\label{rms}}
\end{figure}

\begin{figure}
\plotone{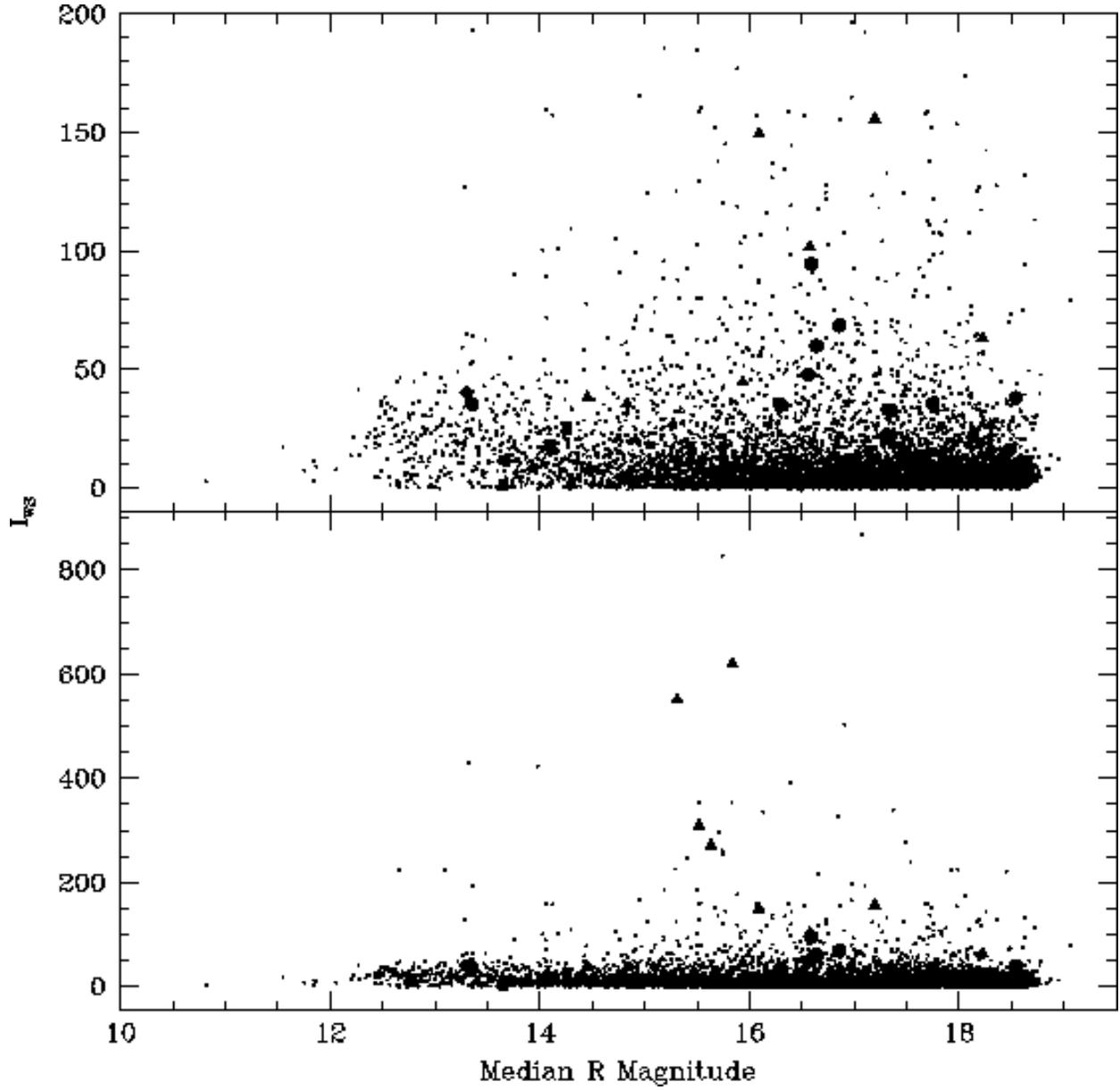}
\caption{Results of the calculation of the Welch-Stetson variability
  index $I_{WS}$ as a function of median $R$ magnitude for the total
  data set.  The top panel is identical to the bottom, highlighting
  the low scoring variables.  The detected variables are denoted as
  follows: \textit{Filled Square}-\ds or pulsating variable,
  \textit{Filled Circle}-W UMa variable, \textit{Filled
  Triangle}-detached eclipsing binary, and \textit{Filled
  Diamond}-unknown/unidentified variable. \label{iws}}
\end{figure}

The search for variable stars in the data set was conducted using
several techniques. Because transits of extrasolar planets exhibit
characteristics different than typical eclipsing binary systems and
pulsating variables (primarily in amplitude and variability
signature), we optimize our detection methods to search for detached
eclipsing binary systems, W UMa variables and \ds and pulsating
variables (that is, the higher-amplitude variables).  Algorithms
optimized for planetary transit searches and low-amplitude variables
will be implemented in a subsequent study.  As a general test of
variability, the RMS variation about the $R$ band median magnitude was
calculated for each star.  The results of this calculation are shown
as a function of median $R$ magnitude in Figure~\ref{rms}.  Because
this index is sensitive to outlier or spurious measurements the
variability statistics $I_{WS}$ from \citet{ws93} and $J$ from
\citet{pbs96} were also employed, the statistics very often used when
searching light curves for variability \citep{he04,br03,mo02}.  These
indices use the correlation between closely-spaced (in time) data
pairs to determine variability.  More specifically, the $I_{WS}$
statistic is given as
\begin{displaymath}
I_{WS}=\frac{1}{\sqrt{n(n-1)}}\sum_{i=1}^{n}\left((\frac{m_1-m_{med}}{\sigma_1})(\frac{m_2-m_{med}}{\sigma_2})\right)_i,
\end{displaymath}
where $m_{1,2}$ are the first and second magnitude measurements of the
$i$th data pair, $\sigma_{1,2}$ are the propogated errors of the
magnitude differences, $m_{med}$ is the calculated median magnitude,
and $n$ is the total number of data pairs.  The results for the
$I_{WS}$ calculation are shown in Figure~\ref{iws}.  Also, the
Lomb-Scargle power spectrum \citep{sc82} was calculated between 1 h
and 1 d (period sampling of 0.001 d) for every light curve in the data
set, and the peak power and corresponding period were recorded.  The
Lomb-Scargle routine can be used in this way as a variability
detection method.  Being a Fourier analysis method, it is particularly
sensitive to stars with periodically varying light curves resembling
sine or cosine functions (such as the W UMa variables).

\begin{figure}
\plotone{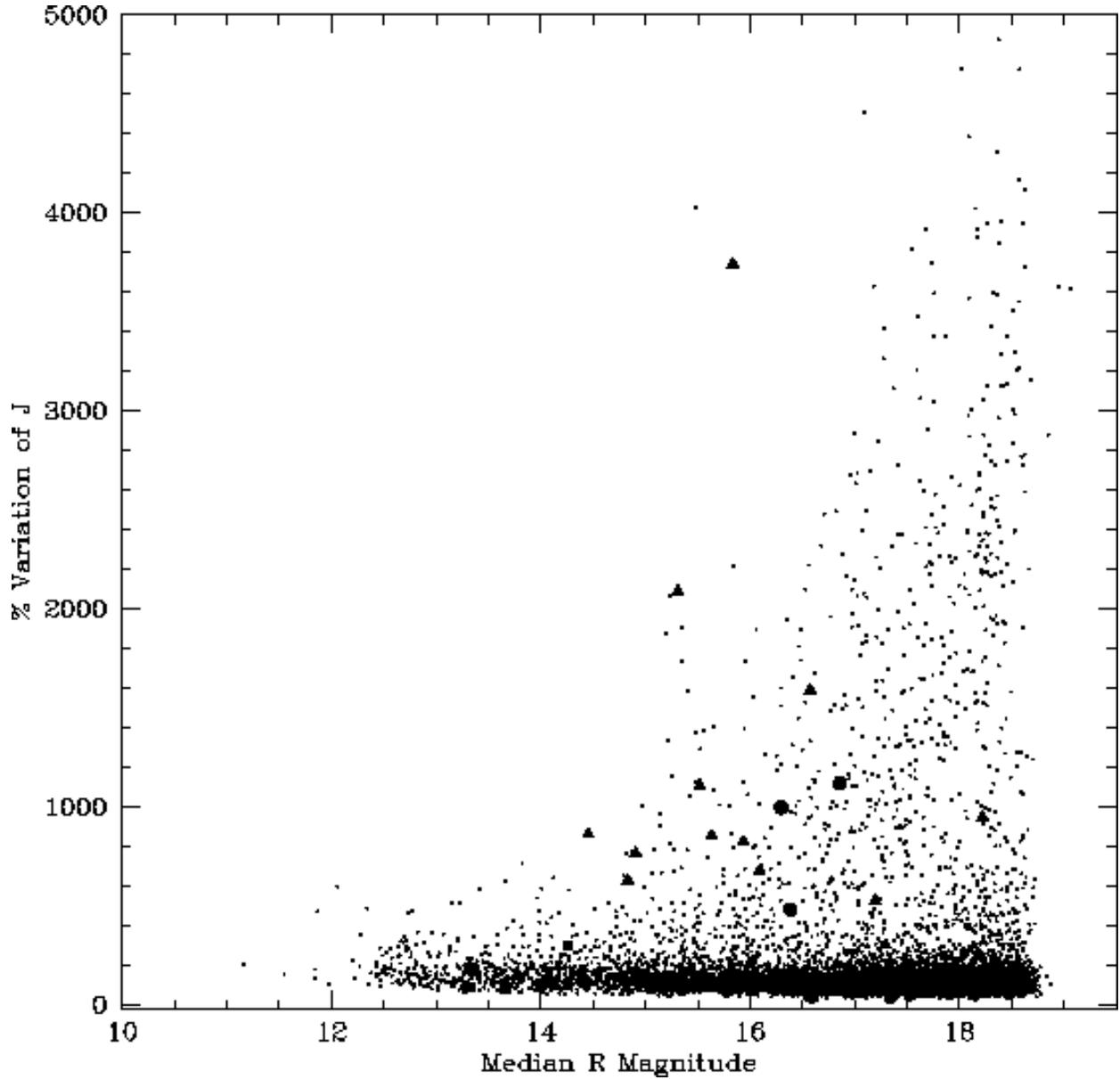}
\caption{Results of the calculation of the percentage of
  night-to-night changes in the Stetson $J$ variability statistic as a
  function of median $R$ magnitude for the total data set. This
  calculation will be sensitive to the detection of eclipsing binary
  stars; see $\S$\ref{detection} for details.  The detected variables
  are denoted as follows: \textit{Filled Square}-\ds or pulsating
  variable, \textit{Filled Circle}-W UMa variable, \textit{Filled
  Triangle}-detached eclipsing binary, and \textit{Filled
  Diamond}-unknown/unidentified variable. \label{pct}}
\end{figure}

While the $I_{WS}$ and $J$ variability statistics are generally more
sensitive to variability than the RMS search, the relatively high
level of the noise in Figure~\ref{iws} makes the detection of
variables difficult. This arises, most likely, as a consequence of
correlated seeing variations in a spatially crowded field when stars
of comparable magnitude have a significant amount of overlap (see
\citealt{sa03b} and \citealt{he04} for further discussion).  To
overcome this difficulty we calculate the $I_{WS}$ and $J$ statistics
for each star \textit{for each night} of data and examine those stars
that have a large percentage change in the statistic from
night-to-night.  This will be particularly useful for detecting
detached eclipsing binary systems, as the system will have little or
no intrinsic variability on some nights (low $I_{WS}$ or $J$ when the
system is not eclipsing) but will have relatively high variability
scores (high $I_{WS}$ or $J$ when the system is eclipsing) on others.
The results of this calculation are displayed in Figure~\ref{pct},
showing that 13 of the 14 detected detached eclipsing binary systems
stand out easily above the noise.  While the $I_{WS}$ and $J$
statistics are imperfect due to the crowded-field conditions, the use
of multiple detection techniques (designed to identify specific types
of variability) significantly increases the probability of detection.
In general, variables were detected by viewing the light curves of
stars above a reasonably low threshold in these statistics
($I_{WS}>15$, $J>15$, percentage variation of $J>200\%$).

Once a variable star was detected, the period was determined using two
slightly different techniques.  Both the Lomb-Scargle \citep{sc82}
periodogram and the variation of the Lafler-Kinman algorithm
\citep{la65} by Deeming \citep{bo70} were calculated for each variable
star that showed multiple events in the data set. The Lomb-Scargle
calculation consists of the fitting of sine and cosine terms of
various frequencies corresponding to possible periodicities in the
data.  The Lafler-Kinman search examines the light curve, folded on a
trial period, for the minimization of differences between data points
that are adjacent in phase space.  Agreement between the two methods
was excellent.

In order to confirm any suspected variable stars that may be in the
M11 field and cross-reference our detections with previous studies, a
search of the General Catalog of Variable Stars \citep{gcvs4} was
undertaken. This search revealed three suspected variables in cluster
field: NSV 11410 (suspected long-period, slow irregular pulsating
variable), 11402 (suspected long-period, slow pulsating variable), and
24615 (unspecified variable type).  The star NSV 24615 is located
near, if not coincident with, the brightest star in the cluster (HD
174512), which was saturated in our study.  Searches around the
coordinates of NSV 11410 and NSV 11402 did find nearby stars but no
variability was detected.  Given that the suspected variables are
proposed slow variables, it remains plausible that the stars actually
are variable in nature but that variability would be undetected over a
short span of observations (3-10 days).  No long-term (year-to-year)
variability was noted in the light curves.  Of the 6 previously known
variable stars (see $\S\ref{known}$), 3 were saturated (V369 Sct and
the 2 spectroscopic binaries of \citet{le89}) and 2 were located
outside the field of study (BS Sct and IT Sct).  Thus, photometry was
obtained only for the star ID \#770 from \citealt{mps77}.  However, no
sign of variability was detected.  The star does appear to be a
cluster member (by position on the CMD) but does not lie in the
theoretical instability strip.  Given that V369 Scuti (and possibly
\#770 from \citealt{mps77}) is the only known variable star in the
field of M11 covered in this study, all the variable stars in this
study are new discoveries.

\section{Color-Magnitude Diagram\label{CMD}}

\begin{figure}
\plotone{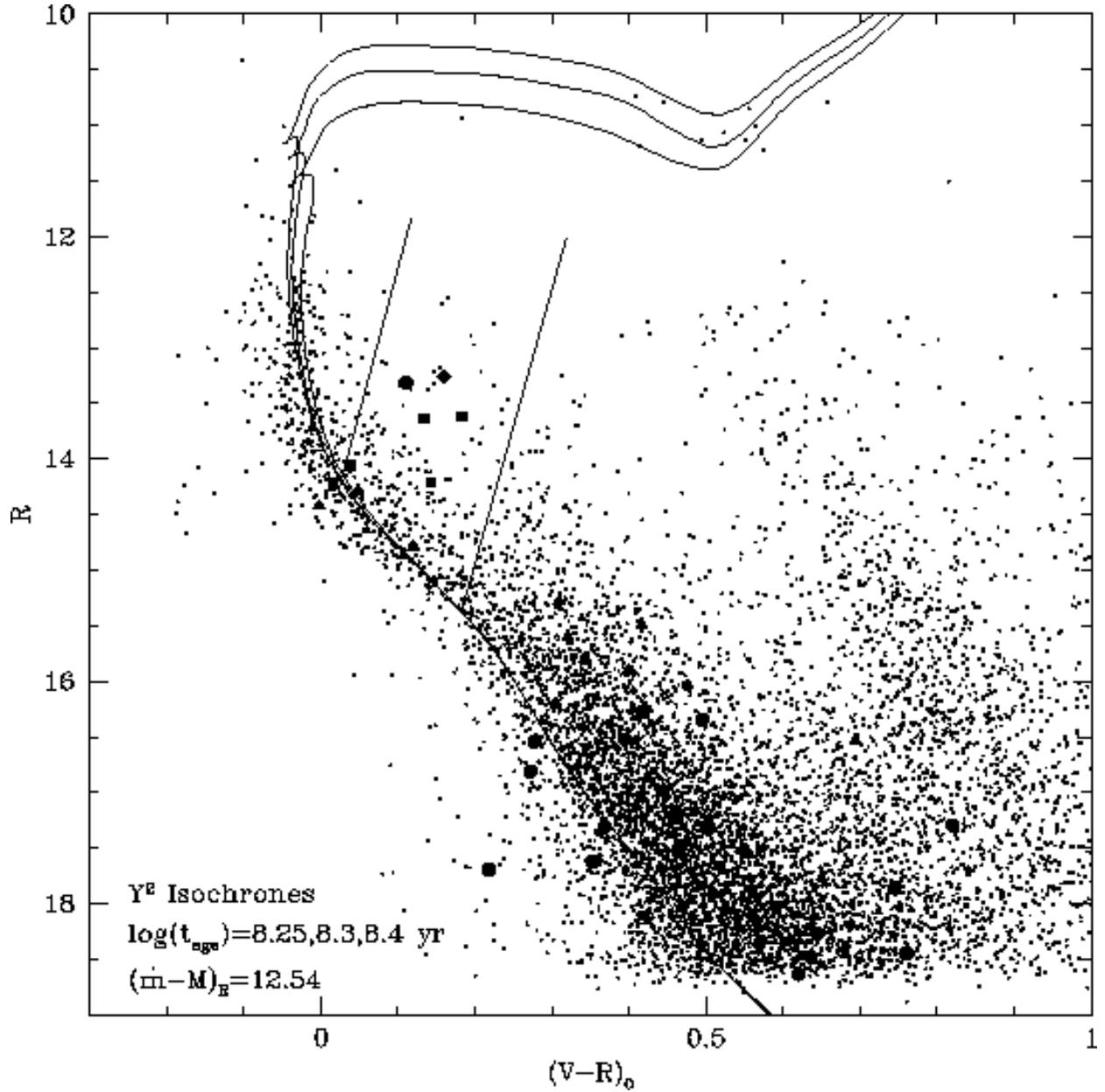}
\caption{$R,(V-R)$ color-magnitude diagram. Overlaid are the Y$^2$
  theoretical isochrones and the theoretical instability strip from
  \citet{pa00}.  The detected variable stars are denoted as follows:
  \textit{Filled Square}-\ds or pulsating variable, \textit{Filled
  Circle}-W UMa variable, \textit{Filled Triangle}-detached eclipsing
  binary, and \textit{Filled Diamond}-unknown/unidentified
  variable.\label{cmdr}}
\end{figure}

\begin{figure}
\plotone{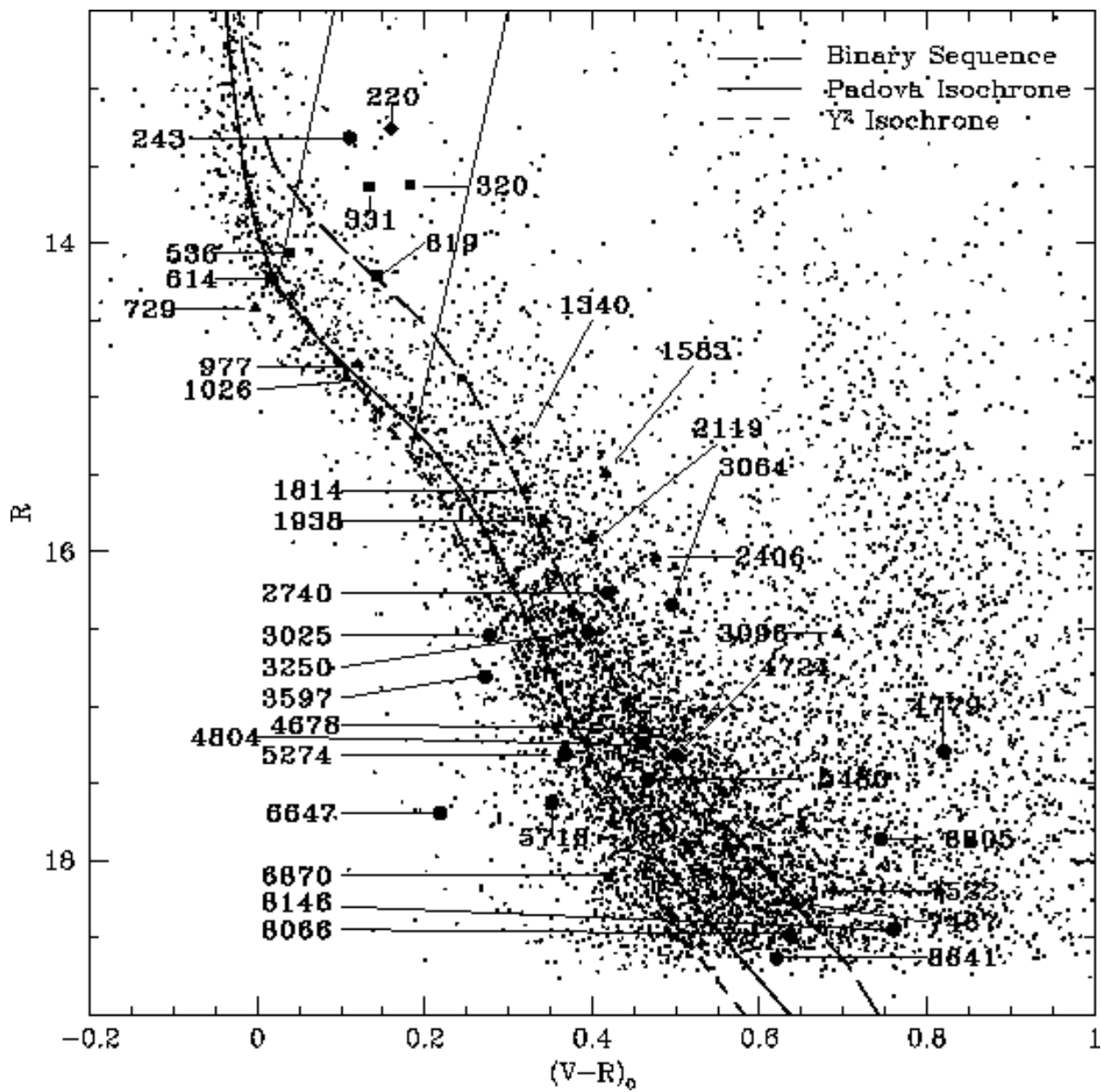}
\caption{$R,(V-R)$ color-magnitude diagram with variable stars
identified. Overlaid are the theoretical isochrone from the Padova and
Y$^2$ groups, the theoretical binary sequence, and theoretical
instability strip from \citet{pa00}.  The symbols are the same as in
Fig. \ref{cmdr}.\label{cmdr_vars}}
\end{figure}

One tool for estimating cluster membership for the detected variable
stars is the cluster color-magnitude diagram (CMD).
Figures~\ref{cmdr} and~\ref{cmdr_vars} show the CMD of the M11 field
derived from the measurement of 8,259 stars in this study; this is the
total sample of stars measured in both the $R$ and $V$ images.  Also
marked are the detected variable stars from Table~\ref{data} and the
theoretical instability strip from \citet{pa00}.  The theoretical
isochrones from the Yonsei-Yale (Y$^2$) stellar evolution models
\citep{ki02} are shown for comparison (color-T$_{\textrm{eff}}$
transformations from \citealt{le98}).  The metallicity was chosen as
$Z=0.02$ in order to make a direct comparison to the CMDs of
\citet{su99}.  Figure~\ref{cmdr_vars} shows the theoretical isochrone
from the Padova group (\citealt{gi00} with color-T$_{\textrm{eff}}$
transformations described in \citealt{gi02}), generated for an
identical metallicity and age ($\log t_{age}=8.3$ yr).  Because the
observed data are uncalibrated, the magnitudes and colors were shifted
to match the theoretical isochrones.
The derived unreddened distance modulus from \citet{su99} was used to
shift the Y$^2$ models to the apparent magnitude scale by adding the
appropriate amount of reddening, assuming $E(B-V)=0.428$ and $R=3.1$
\citep{su99}.  This shifting of the observed data was done only for
convienent plotting.  We have not produced standard magnitudes, and we
have not done anything requiring standard magnitudes.

As first noted by \citet{ma84}, the CMD of the M11 field appears to
have not only the typical red field star population, which is easily
distinguished from the cluster main-sequence (MS), but also has a
population of stars overlapping the cluster MS starting at
approximately $V\sim15$ ($R\sim15.5$) and fainter. The
field population is bimodal; there exists a ``red'' field star
population ($(V-R)_{o}>0.7$) and a ``blue'' field star population
($0.25<(V-R)_{o}<0.7$).  The blue field star population
hampers the identification of cluster members via the CMD for many of the
detected variable stars in this study.  Membership for individual
variables will be discussed in $\S\ref{ds}-\ref{wumas}$.

\section{Light Curves\label{lcs}}

\subsection{\ds Variables\label{ds}}

\begin{figure}
\plotone{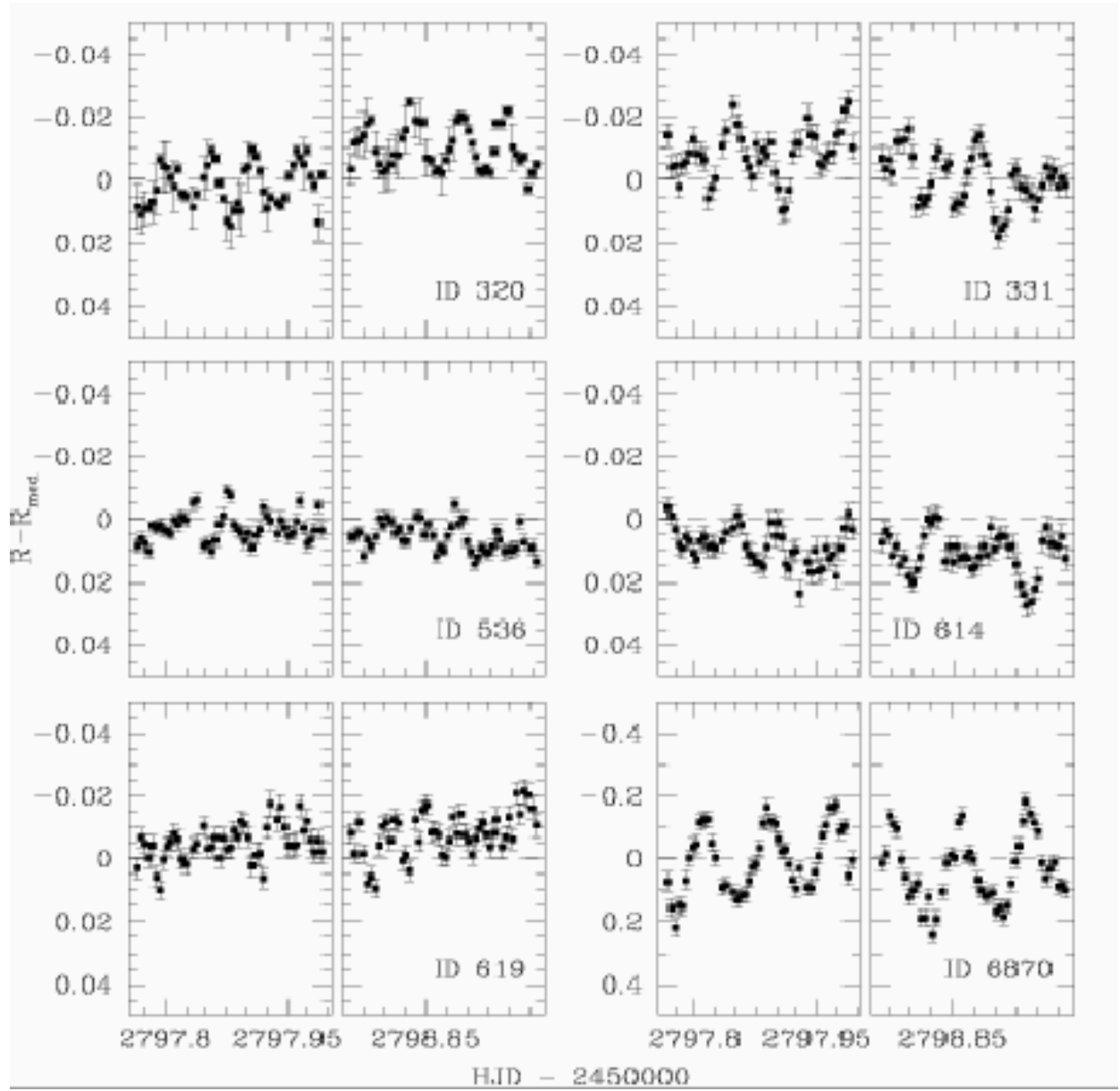}
\caption{Light curves for the six \ds variables detected in this
 study.  Two nights of data are shown for each star.\label{phase_ds}}
\end{figure}

In Figure~\ref{phase_ds}, we show sample light curves for the 6 \ds
variables (IDs 320, 331, 536, 614, 619, and 6870) detected in this
study.  As can be seen in Figures~\ref{cmdr} and~\ref{cmdr_vars}, the
cluster MS just begins intersects the faint edges of the theoretical
instability strip \citep{pa00} and so a \ds population in this cluster
is not surprising.  Given the bright magnitude of these objects,
membership probabilities are available from the \citet{mps77} proper
motion study; these are given in Table~\ref{data}.

Particularly interesting is the variable ID 6870, which shows an
amplitude of variation more than an order of magnitude greater than
the five other detected \ds variables.  This star is likely a member
of a sub-class of \ds variables known as high amplitude \ds variable
stars (see \citealt{ro04} and references therein).  These differ from
regular \ds pulsators in that they have significantly larger
amplitudes of variation (often greater than 0.3 magnitudes), appear to
have only one or two radial pulsation modes (and no non-radial modes),
and have slower rotation speeds \citep{ro96}.  ID 6870 is likely a
field population main-sequence star, as \ds variability would position
the star inside or near the instability strip if it were a cluster
member.  Use of the period-density relation for \ds stars \citep{br00}
yields a stellar density typical of MS stars.

\subsection{Semi-detached and Detached Eclipsing Binaries\label{eb}}

\begin{figure}
\plotone{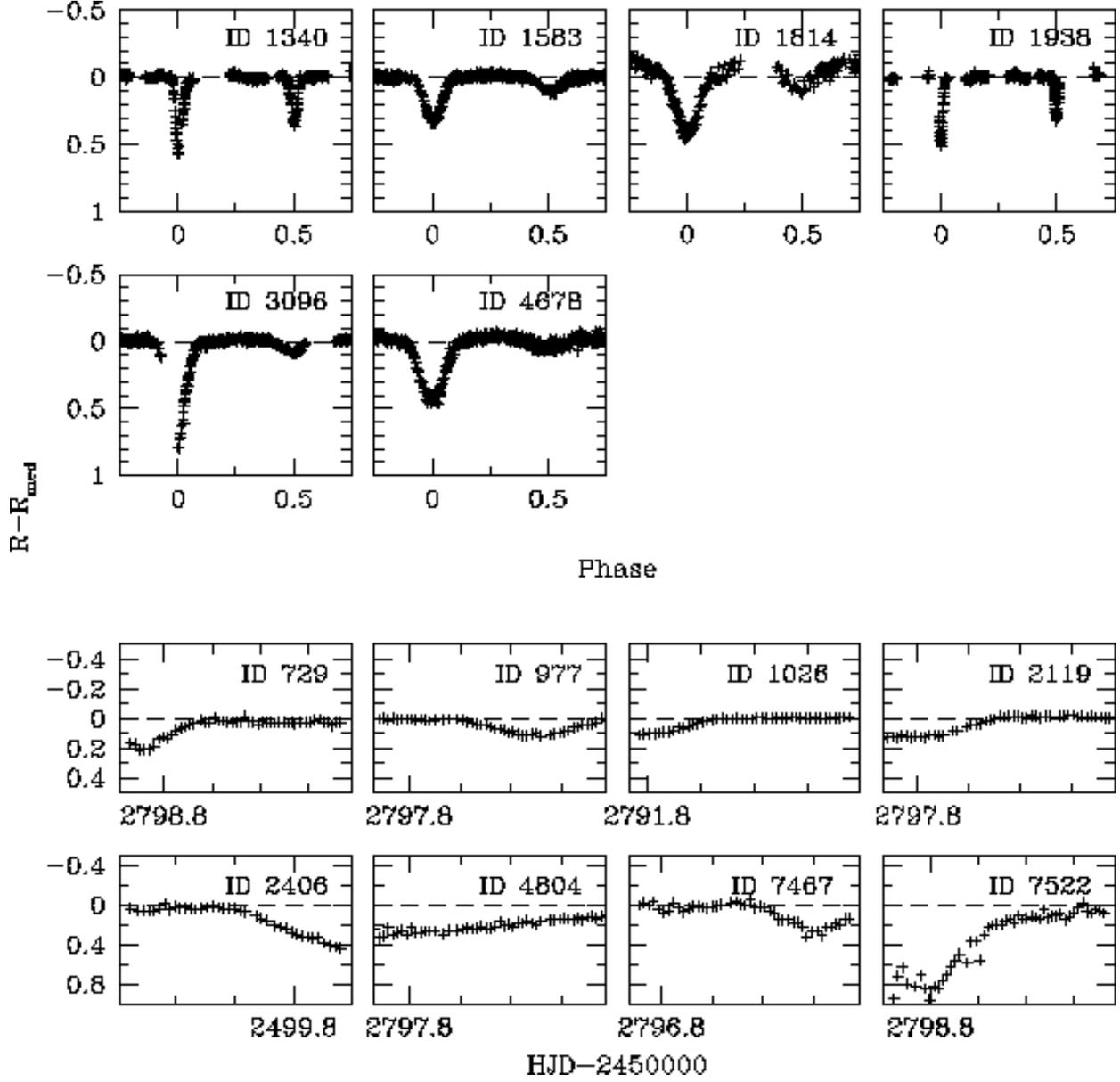}
\caption{Light curves (types EA or EB) for eclipsing binary systems
  detected in this study.  The top panels show those eclipsing
  binaries where multiple eclipses were observed, allowing for the
  determination of an ephemeris.  The bottom panels shows eclipse
  events for those systems where only one or two eclipses were
  detected and no ephemeris could be determined.\label{phase_deb}}
\end{figure}

In Figure~\ref{phase_deb} we show the light curves for the 14 detached
eclipsing binary systems detected in this study.  Multiple eclipses
were detected for 6 of the 13 systems, and hence periods were
determined (listed in Table~\ref{data}) for these systems. IDs 1814
and 4678 show characteristics of RS CVn-type eclipsing binaries,
namely the large light curve scatter and rapidly changing light curve
shapes (presumably due to quickly changing star spot activity).  IDs
1583 and 3096 show light curves typical of Algol-type semi-detached
systems while IDs 1340 and 1938 show little or no outside of eclipse
light variations and are likely detached systems. In 8 of the 14
detected systems only one or two events were detected and therefore
periods could not be determined.

In terms of cluster membership, we are only able to make reasonable
judgments about the brightest stars owing to the overlap of the field
star population noted in $\S$\ref{CMD}.  Both IDs 729 and 977 show
positions on the CMD consistent with cluster membership.  The proper
motion study by \citet{mps77} further supports this conclusion where
they determine membership probabilities of $81\%$ and $52\%$, for IDs
729 and 977, respectively.  While the \citet{mps77} finds a very low
membership probability for ID 1026 ($7\%$), the position of this
system near the cluster MS suggests it is likely a cluster member. The
stars ID 1340, 1814, and 1938 show positions near the equal mass
binary sequence, but given the presence of the overlapping field star
population below $R=\sim15.5$, we are unable to definitively claim
cluster membership without further data on these or fainter systems.
Finally, although ID 2119 shows total eclipses in its light curve, its
position in the CMD indicates that it is almost certainly a part of
the field population, so we have not pursued it further.

\subsection{W Ursae Majoris Variables\label{wumas}}

\begin{figure}
\plotone{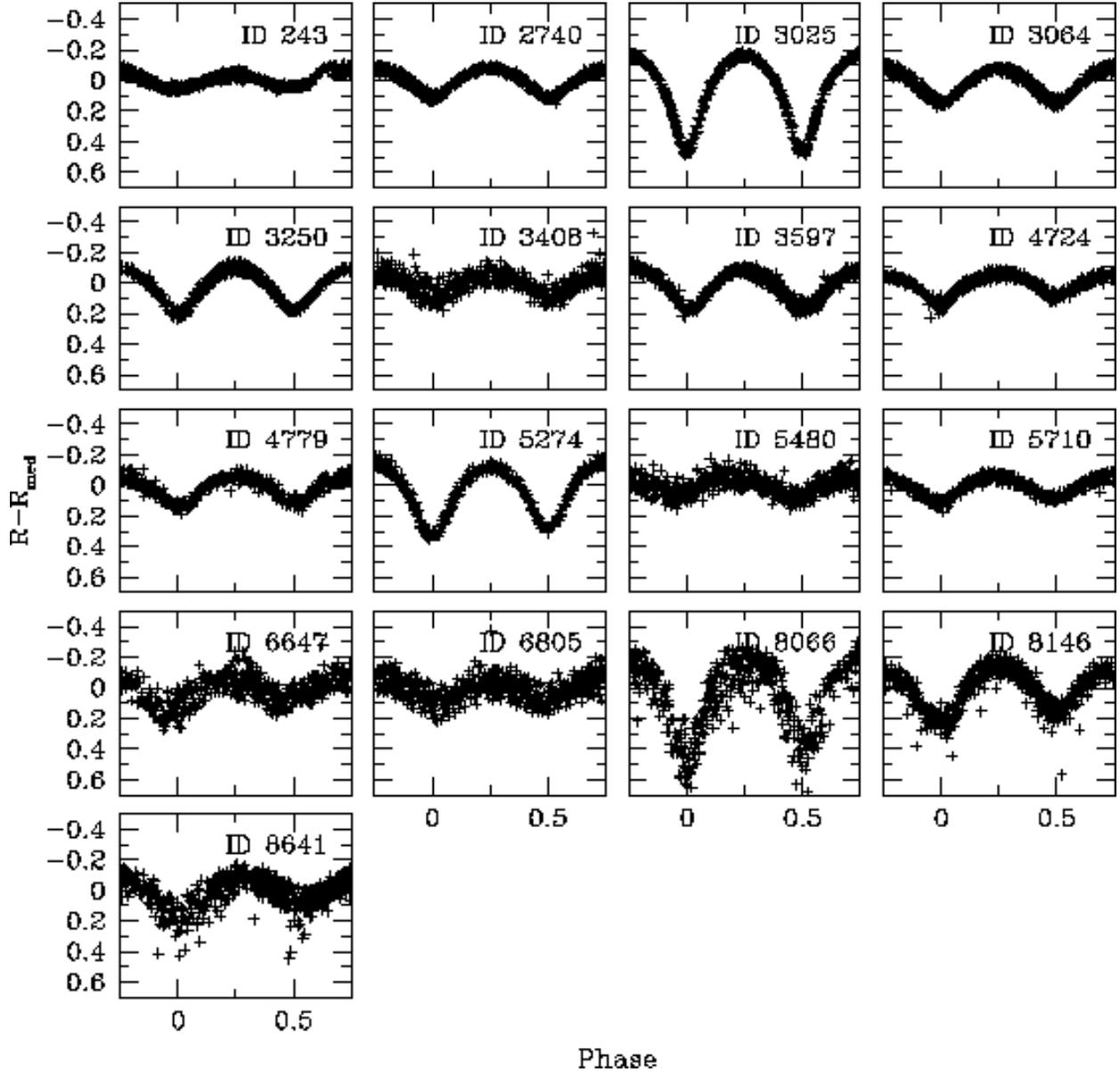}
\caption{Light curves (type EW) for the W UMa variable stars detected
  in this study.\label{phase_wumas}}
\end{figure}

In Figure~\ref{phase_wumas} we show the light curves for the 17 W UMa
variables (contact binary systems) detected in this study.  The
differences in shapes of the light curves of these systems are
primarily due to three main effects: the orbital inclination
(influencing the depth of eclipses), the degree of fillout of the
Roche geometry (influencing the steepness of the eclipses and the
``broadness'' of the light curves at quadratures), and the mass ratio.
In order to make a judgment regarding the geometric configuration of
these systems, we modeled the data of the brightest systems using the
\textit{Binary Maker 3} light curve synthesis program \citep{br02}.
Given the lack of spectroscopically determined mass ratios, elaborate
fitting of the data using the Wilson-Devinney code (\citealt{wd71},
hereafter WD) or other fitting routines is not worthwhile for those
systems that do not exhibit total eclipses. A mass ratio of $q=0.4$
was assumed when modeling those systems.  Modeling shows that these
variables can easily be described as contact binary systems.  We
discuss the light curves and synthetic models of IDs 3025 and 5274
below, both of which show evidence for total eclipses. In the case of
totally-eclipsing light curves the assumption of an arbitrary mass
ratio is unnecessary; the duration of totality strongly constrains the
possible combinations of inclination and mass ratio.

\begin{figure}
\plotone{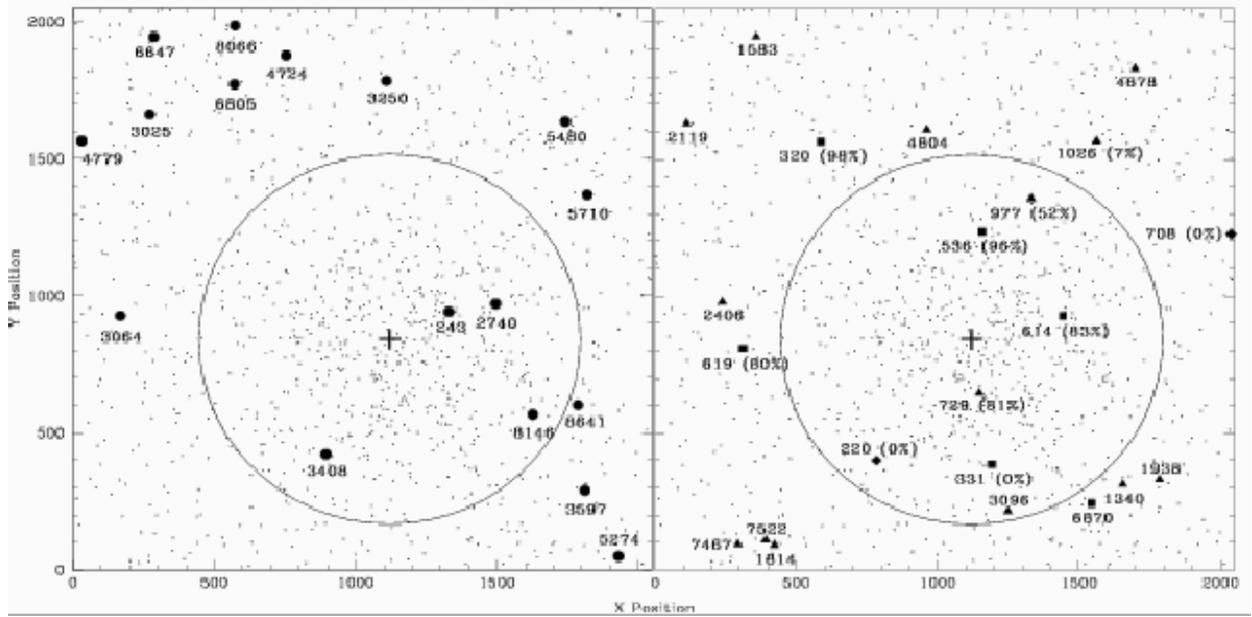}
\caption{Reference frame positions for the variables detected in this
study.  The left panel shows the W UMa variables (\textit{Filled
Circles}) while the right panel shows the detected detached eclipsing
binary systems (\textit{Filled Triangles}), \ds or pulsating variables
(\textit{Filled Square}), and unknown/unidentified variables
(\textit{Filled Diamond}). Shown for reference are stars brighter than $R=15$
(\textit{Small Points}) and the cluster half-mass radius ($r=$4\Min5;
\citealt{ma84}).\label{spatial_both}}
\end{figure}

The CMD presents the primary tool by which to estimate cluster
membership for these contact systems. It is likely that the detected
systems are members of the field population for two reasons: (1) the
large spatial separation of systems from the cluster center and (2)
positions on the CMD which are inconsistent with the cluster MS and
binary sequence.  In the first case, we find that the radial distances
from cluster center (column 5, Table~\ref{data_pa}) are likely
inconsistent with cluster membership in all but five (IDs 243, 2740,
3408, 8146, 8641) of the 17 cases.  Figure~\ref{spatial_both} shows
the positions of the detected W UMa systems relative to the brightest
non-varying cluster stars ($R>15$), the cluster center, and the
cluster half-mass radius (as determined by \citealt{ma84}).  For
reference, the adjacent panel shows the positions of the other
detected variables.  These indicate that the W UMa systems are more
concentrated within the field population than the other 3 types of
detected variables, especially considering that dynamical evolution
should concentrate member binary systems to the cluster center.  In
the second case, only 3 of the 17 W UMa systems (IDs 5480, 8066, 8641)
have CMD positions between the cluster MS and binary sequence.  While
this could be indicative of cluster membership, we stress that the
possible presence of differential reddening, statistically uncertain
$V$ magnitudes that contribute to the $(V-R)$ position on the CMD, and
the overlapping field star population in the CMD complicate these
judgments.  In summary, given the two criteria above for consideration
of cluster membership, only ID 8641 appears to have both a radial
distance from cluster center and CMD position consistent with
membership.  While ID 2740 is located spatially close to cluster
center, it is not easily distinguishable from the field star
population on the CMD.  No $V$ measurement was obtained for ID 3408
due to the crowded nature of the M11 field, and hence no CMD
information is available for that system.  Because proper motion data
is available for ID 243, we discuss the possibility of cluster
membership below.

\begin{figure}
\plotone{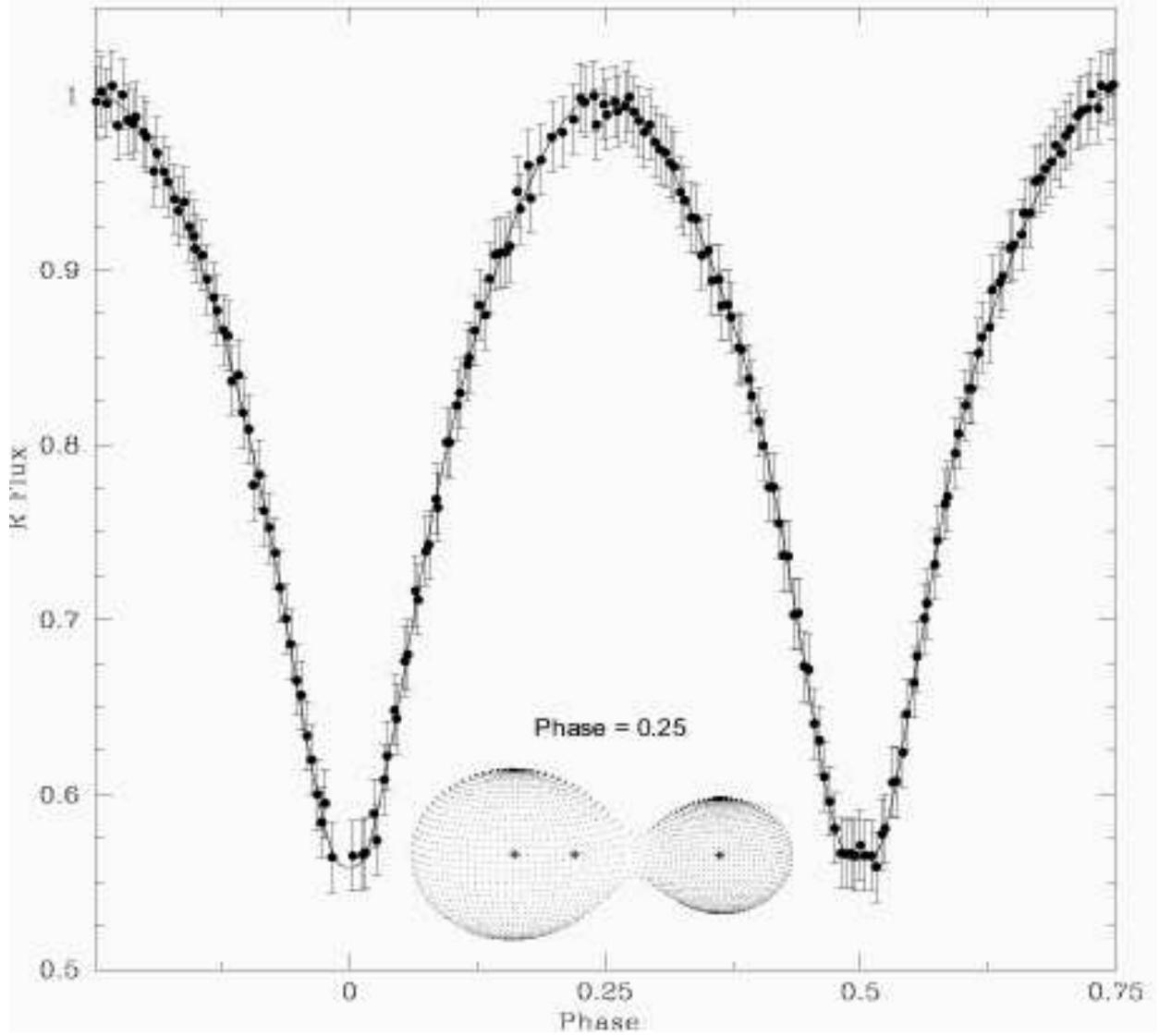}
\caption{Phased light curve of ID 3025 (200 \textit{R}-band normal points)
  shown with the adopted Wilson-Devinney differential corrections
  synthetic fit.\label{3025}}
\end{figure}

\noindent\textbf{ID 3025} Short total eclipses are evident in the
light curve of the variable ID 3025 in Fig. \ref{3025}.  Initial
models were generated using \textit{Binary Maker 3} and the WD code
was used to refine these input model parameters. Figure~\ref{3025}
shows the observed light curve of ID 3025 (data have been binned into
200 normal points), the 3-D model generated from \textit{Binary Maker
3}, and the WD solution.  The model parameters (adopted and fit) are
given in Table~\ref{3025_table}.  The ephemeris (probable errors in
parentheses) derived from the period study is: time of primary
minimum=2452495.74767+0.441864(3). Our analysis indicates that ID 3025
is likely a typical A-type contact system, the larger star being
hotter by approximately 100 K.

\begin{figure}
\plotone{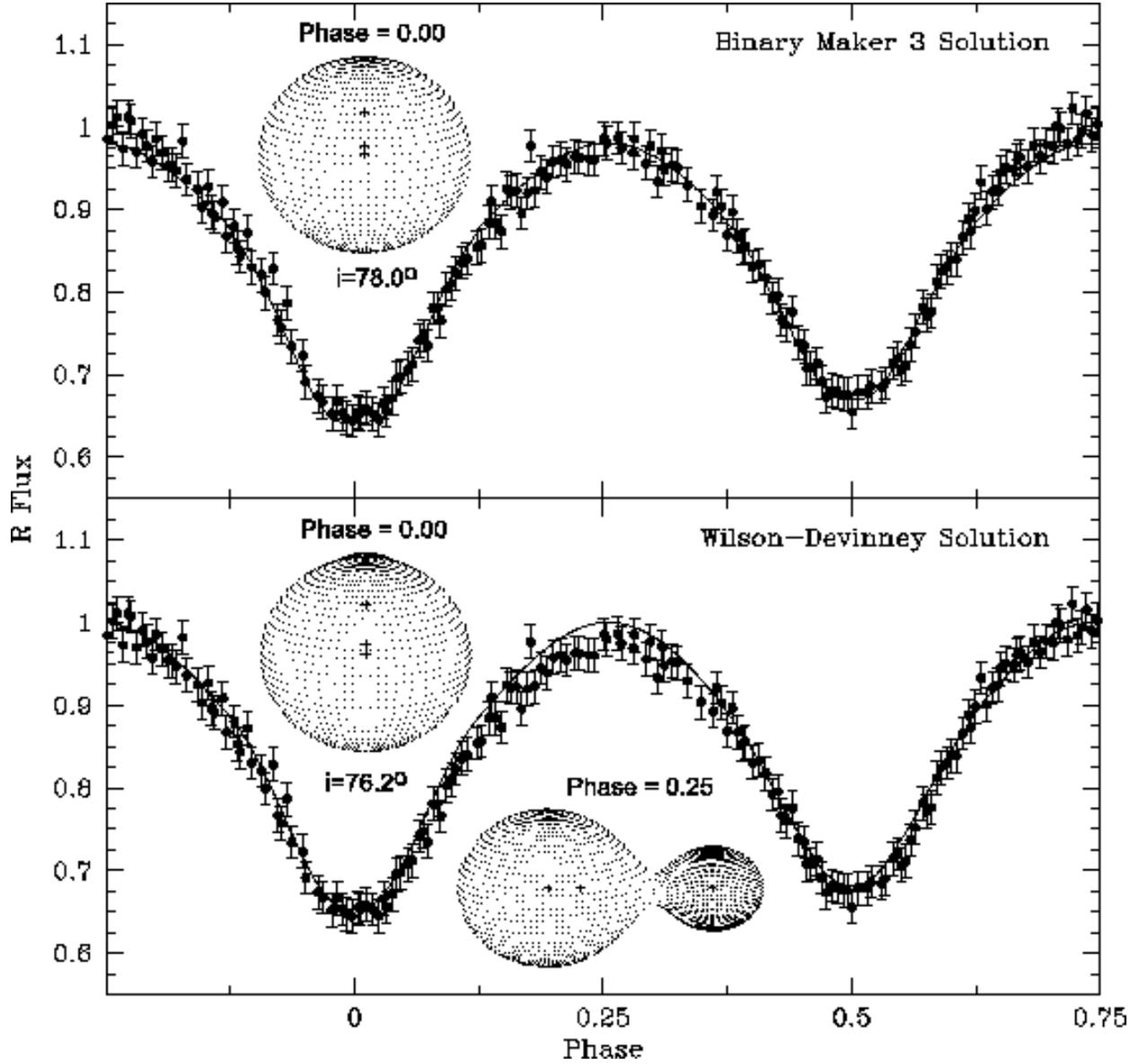}
\caption{Phased light curve of ID 5274 (200 \textit{R}-band normal
  points) shown with the (\textit{top panel}) initial \textit{Binary
  Maker 3} synthetic fit and (\textit{bottom panel}) adopted
  Wilson-Devinney differential corrections synthetic fit.  The 3-D
  models are displayed at phase 0.00 in both cases, showing that the
  Wilson-Devinney model just barely produces a non-total primary
  eclipse.\label{5274}}
\end{figure}

\noindent\textbf{ID 5274} Individual nights of data for the variable ID
5274 show evidence of a total primary eclipse.  An initial model was
generated using \textit{Binary Maker 3}, biasing the solution to one
with a combination of mass ratio, fillout and inclination that matched
the total eclipse.  To take a more unbiased approach to the analysis
we used this initial model as input to WD to derive a final
differential corrections solution.  Figure~\ref{5274} shows the
observed light curve of ID 5274 (data have been binned into 200 normal
points), along with these two solutions.  The model parameters are
given in Table~\ref{5274_table}. We find the WD solution highly
influenced by (1) the large scatter in the light curve and (2) the
significant asymmetry in the quadratures (likely due to the presence
of cool spots). The WD corrections appear to be favoring a lower
inclination solution in order to fit the shoulders of quadrature
around phase 0.25.  Thus, the WD solution represents an averaging of
the two different brightnesses at quadrature. We did not model the
presence of spots in this system because of the large scatter and lack
of multibandpass observations. Our analysis finds ID 5274 to be a
W-type contact binary system, the larger star being cooler by
approximately 450 K. The ephemeris (probable errors in parentheses)
derived from the period study is: time of primary
minimum=2452796.92403+0.3468800(1). While we believe the system to be
totally eclipsing, the lower inclination solution calls into question
the derived mass ratio.  Given the large scatter in the data and the
uncertainties in the inclination/mass ratio combination, these models
should be considered preliminary.  Higher precision, multifilter light
curves are necessary to further constrain these models.

\noindent\textbf{ID 243} For the variable ID 243, independent cluster
membership information is available from the proper motion study of
\citet{mps77}, who find a moderate probability of membership ($62\%$).
This object is also radially located within the cluster half-mass
radius adding further circumstantial evidence that this variable could
be a cluster member.  In terms of location on the CMD, this object
falls inside the theoretical instability strip.  It is unlikely,
however, that this object is a cluster pulsating $\delta$ Scuti
variable since the necessary period ($P=0.43289$ d for 1 pulse per
cycle) would be at least an order of magnitude greater than typical
$\delta$ Scuti variables.  The light curve of ID 243 is consistent
with those of very low inclination W UMa variables, and given the
longer period of the system, this presents a more reasonable
interpretation of the variability.  Spectroscopic data would help in
the resolution of the true nature of this system and refine the
question of membership in M11.

Despite this circumstantial evidence for cluster membership, the young
age of the cluster may preclude a population of W UMa members.  The
presence of contact binary stars in young clusters would likely
require either (1) the formation of systems with small initial orbital
separations, such that an angular momentum loss mechanism could bring
the stars into contact on a short timescale \citep{gu88}, or (2) the
birth of binary systems in the contact phase.  While these scenarios
may not be unlikely, the detection of contact systems in young or
intermediate age open clusters (ages less than $\sim1$ Gyr) remains
elusive.  Observational difficulties result from the presence of faint
W UMa variables in a bright, crowded field; image exposure times must
be long to detect faint variables, but longer exposure times result in
over-exposure of the brightest stars.

\subsection{Candidate Variables\label{canidate}}

\begin{figure}
\plotone{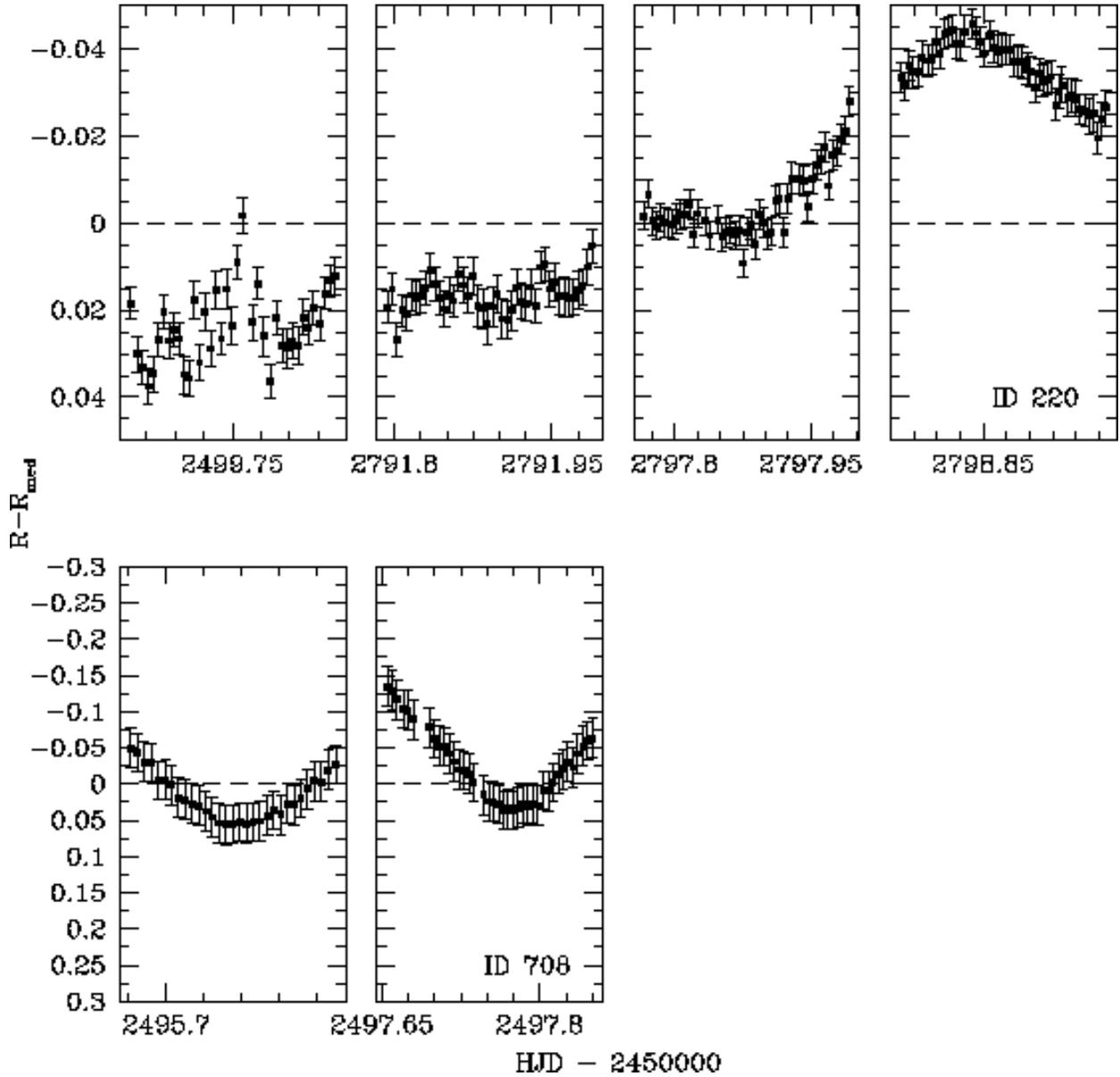}
\caption{Light curves for the unknown/unidentified variables detected
in this study.  Four nights of data are shown for ID 220 (showing both
the long and short term variations).  Two nights of data are shown for
ID 708.\label{unknown}}
\end{figure}

In this study, two variable stars were detected that could not be
readily identified from the shape of the light curves.  The data are
shown in Figure~\ref{unknown} and we describe them individually below.

\noindent\textbf{ID 220} This object, while located within the
theoretical instability strip on the cluster CMD (see
Figure~\ref{cmdr_vars}), has a membership probability of 0\% as
determined by \citet{mps77}.  However, the proper motion study by
\citet{su98} finds a membership probability of 100\%.  Photometric
variations are observed on both long (as shown in
Figure~\ref{unknown}) and short timescales (at nearly the Nyquist
frequency of the observations, $\sim10$ min).  The presence of the
short or long term variability was observed to differ from
night-to-night; thus, it is unclear to which category of variable
stars this object belongs. The appearance of multiple frequencies of
variability and possible cluster membership (and hence a CMD position
in the instability strip) tends to argue for a type of pulsating
variable.

\noindent\textbf{ID 708} This object was near the edge of the
reference frame and hence was only measured in two nights of
observing.  No $V$ band data were obtained for this star (since it was
located just outside the reference frame), so the position on the CMD
was not determined.  It is unclear if the minima in the light curves
are a result of eclipses (from either a detached eclipsing binary or
contact binary system) or minimums in stellar pulsation.

\section{Conclusions\label{conc}}

In conclusion, this work presents the first large-scale photometric
variability survey of the intermediate age ($\sim200$ Myr) open
cluster M11.  A total of 39 variable stars were detected and can be
categorized as follows: 1 irregular (probably pulsating) variable, 6
\ds variables, 14 detached eclipsing binary systems, 17 W UMa
variables, and 1 unidentified/candidate variable.  While previous
proper motion studies \citep{mps77,su98} allow for cluster membership
determination for the brightest stars, we find that the membership
determination is significantly hampered below $V=15,R=15.5$ by the
large population of field stars at similar photometric colors to the
cluster MS.

Of the detected systems, several are worth noting for follow-up work.
First, three detached eclipsing binaries (IDs 729, 977, 1026) are
likely cluster members, but will require further photometric data to
deduce the orbital period.  Given more photometric and spectroscopic
data on these systems, the derivation of the absolute parameters
(masses and radii) could be useful in constraining theoretical stellar
models.  Second, of the six \ds variables detected in this study (all
of which are likely cluster members), further studies of the pulsation
periods of IDs 536 and 614 would make excellent tests of the blue edge
of the theoretical instability strip.  Third, the W UMa variable ID
243 is a potential cluster member.  If it can be confirmed that it is
a cluster member and also a W UMa variable, it would be the first
confirmation of a contact system in an open cluster younger than 0.7
Gyr.  Further photometric study can help confirm the orbital period
and light curve variations, while spectroscopic observations will be
critical to confirming the binary nature of the system.  If the system
is in fact a contact binary, it is likely that the spectrum will show
double lines.  Lastly, we find two W UMa variables that exhibit total
eclipses in their light curves and present preliminary models using
\textit{Binary Maker 3} and Wilson-Devinney differential corrections.
While the total secondary eclipse of the field W UMa variable ID 3025
constrains the mass ratio of the system photometrically, a
spectroscopic mass ratio would be more appropriate and help in
determination of the absolute parameters of the system.  Inspection of
the light curve of ID 5274 yields evidence of a total primary eclipse.
However, the large scatter in the data and asymmetries in the light
curve favor a lower inclination solution that forces a partial eclipse
model.  Futher high precision, multifilter light curves and a
spectroscopic mass ratio are necessary to further refine our
preliminary models of this system.

\acknowledgments{This research has made use of the SIMBAD database,
operated at CDS, Strasbourg, France.  This project has also made use
of the CALEB database (\url{http://caleb.eastern.edu/}). J.R.H. wishes
to thank Paul Etzel both for useful conversations regarding variable
stars and the generous allocation of observing time at Mt. Laguna
Observatory.  J.R.H. also wishes to thank Slavek Rucinski for useful
conversations about W UMa variables in young open clusters.  The
authors would like to thank the referee for the useful comments which
have greatly improved this paper.  This work has been partially funded
through grant AST 00-98696 from the National Science Foundation to
E.L.S. and Michael Bolte, and through a Research, Scholarship, and
Creative Activity (RSCA) grant from San Diego State University to
E.L.S.}


\begin{deluxetable}{lcc}
\tablewidth{0pt} \tablecaption{Observation Log for M11}
 \tablehead{\colhead{UT Double Date} & \colhead{\# Images} &
\colhead{Average Seeing (\arcsec)}} 
\startdata
08-09 August 2002  & 31 & $2.9\pm0.5$\\
10-11 August 2002  & 40 & $3.8\pm0.6$\\
12-13 August 2002  & 39 & $4.7\pm1.8$\\
13-14 August 2002  & 38 & $5.0\pm1.9$\\
14-15 August 2002  & 37 & $4.5\pm0.9$\\
15-16 August 2002  & 35 & $4.8\pm0.6$\\
30-31 May 2003 &   18 & $4.8\pm0.7$\\
31-01 May 2003 &  45 & $4.9\pm0.8$\\
01-02 June 2003 &  39 & $6.2\pm3.0$\\
02-03 June 2003 &  48 & $6.9\pm3.3$\\
05-06 June 2003 &  44 & $6.4\pm3.3$\\
06-07 June 2003 & 50 & $4.9\pm1.8$\\
07-08 June 2003 &  49 & $4.5\pm0.6$\\
\enddata
\label{obs}
\tablecomments{The photometry in this study is uncalibrated.}
\end{deluxetable}

\begin{deluxetable}{lccccccccccc}
\rotate
\tabletypesize{\scriptsize}
\tablewidth{0pt}
\tablecaption{Detected Variable Stars}
\tablehead{\colhead{ID} & \colhead{GSC,ID$_{\textrm{MPS}}(P\%$)} &
\colhead{$RA$} & \colhead{$DEC$} & \colhead{$m_{V}$} &
\colhead{$\sigma_{V}$} & \colhead{$m_{{R}_{med}}$} &
\colhead{$\sigma_{{R}_{med}}$} & \colhead{$(V-R)_{O}$} &
\colhead{$I_{WS}$} & \colhead{$\sigma_R$} & \colhead{T\tablenotemark{a}}}
\startdata
220 & 683(0) &  18h 51m 14.8s &  $-06\arcdeg18\arcmin26\farcs3$ & 13.665 & 0.006 & 13.2648 & 0.0015 & 0.160 & 45.06 & 0.016 & 4\\
243 & 5126:4687,1268(62)& 18h 51m 00.2s &  $-06\arcdeg14\arcmin 49\farcs2$ & 13.706 & 0.006 & 13.3183 & 0.0050 & 0.110 & 78.94 & 0.041 & 3\\
320 & 5126:5601,1711(98)&  18h 50m 43.6s & $-06\arcdeg19\arcmin47\farcs5$ & 14.044 & 0.006 & 13.6210 & 0.0014 & 0.183 &  5.59 & 0.013  & 1\\
331 & 5126:4823,676(0) & 18h 51m 15.1s &  $-06\arcdeg15\arcmin43\farcs8$ & 14.006 & 0.006 & 13.6330 & 0.0021 & 0.134 & 24.14 & 0.017  & 1\\
536 & 5126:5659,1531(96)& 18h 50m 52.4s & $-06\arcdeg15\arcmin59\farcs1$ & 14.339 & 0.007 & 14.0610 & 0.0021 & 0.038 & 10.15 & 0.016   & 1\\
614 & 5126:5664,1237(83)& 18h 51m 00.6s & $-06\arcdeg14\arcmin04\farcs4$ & 14.479 & 0.007 & 14.2232 & 0.0015 & 0.016 & 38.97 & 0.017   & 1\\
619 &1097(80)&  18h 51m 03.9s &  $-06\arcdeg21\arcmin35\farcs2$ & 14.595 & 0.007 & 14.2124 & 0.0017 & 0.143 & 57.77 & 0.018  & 1\\
708 & 1521(0) & 18h 50m 52.5s & $-06\arcdeg10\arcmin07\farcs7$ & \ldots  & \ldots & 14.4427 & 0.0213 &\ldots & 7.39  & 0.047  & 4\\
729 & 900(81) & 18h 51m 08.1s &  $-06\arcdeg16\arcmin01\farcs7$ & 14.659 & 0.007 & 14.4220 & 0.0022 &-0.003 & 46.98 & 0.040  & 2\\
977 &1596(52)& 18h 50m 49.1s & $-06\arcdeg14\arcmin49\farcs7$ & 15.146 & 0.007 & 14.7862 & 0.0019 & 0.120 & 45.78 & 0.026 &  2\\
1026& 1715(7)& 18h 50m 43.4s & $-06\arcdeg13\arcmin18\farcs0$ & 15.220 & 0.007 & 14.8742 & 0.0016 & 0.106 &38.043 & 0.018 &  2\\
1340& \ldots & 18h 51m 17.0s &  $-06\arcdeg12\arcmin38\farcs6$ & 15.825 & 0.009 & 15.2757 & 0.0017 & 0.309 &224.04 & 0.104 &  2\\
1583& \ldots &  18h 50m 33.4s & $-06\arcdeg21\arcmin19\farcs2$ & 16.149 & 0.009 & 15.4925 & 0.0032 & 0.416 &597.41 & 0.093  &  2\\
1814& \ldots &  18h 51m 23.1s & $-06\arcdeg20\arcmin49\farcs2$ & 16.157 & 0.012 & 15.5967 & 0.0114 & 0.320 &219.03 & 0.208  &  2\\
1938& \ldots & 18h 51m 16.5s & $-06\arcdeg11\arcmin47\farcs3$  & 16.377 & 0.009 & 15.7942 & 0.0022 & 0.343 &604.43 & 0.094  &  2\\
2119& \ldots &  18h 50m 41.9s & $-06\arcdeg22\arcmin56\farcs9$ & 16.543 & 0.009 & 15.9045 & 0.0016 & 0.399 &  7.15 & 0.025  &  2\\
2406& \ldots &  18h 50m 59.4s & $-06\arcdeg22\arcmin05\farcs3$ & 16.750 & 0.010 & 16.0339 & 0.0077 & 0.476 &\ldots & 0.109  &  2\\
2740& \ldots & 18h 50m 59.4s & $-06\arcdeg13\arcmin43\farcs5$ & 16.907 & 0.010 & 16.2690 & 0.0099 & 0.418 & 47.01 & 0.066 &  3\\
3025& \ldots &  18h 50m 41.1s & $-06\arcdeg21\arcmin53\farcs1$ & 16.996 & 0.012 & 16.5437 & 0.0227 & 0.277 &117.17 & 0.219 &  3\\
3064& \ldots &  18h 51m 00.8s & $-06\arcdeg22\arcmin33\farcs6$  & 17.080 & 0.010 & 16.3454 & 0.0098 & 0.495 & 21.24 & 0.079  &  3\\
3096& \ldots & 18h 51m 19.7s & $-06\arcdeg15\arcmin20\farcs8$  & 17.467 & 0.010 & 16.5342 & 0.0022 & 0.693 &739.04 & 0.145  &  2\\
3250& \ldots & 18h 50m 37.7s & $-06\arcdeg16\arcmin19\farcs2$ & 17.056 & 0.010 & 16.5221 & 0.0128 & 0.394 & 59.21 & 0.100  &  3\\
3408& \ldots & 18h 51m 14.2s &  $-06\arcdeg17\arcmin42\farcs2$ & \ldots & \ldots& 16.5916 & 0.0079 &\ldots & 89.16 & 0.094  &  3\\
3597& \ldots & 18h 51m 17.6s &  $-06\arcdeg11\arcmin36\farcs9$ & 17.303 & 0.011 & 16.8114 & 0.0113 & 0.272 & 34.01 & 0.092  &  3\\
4678& \ldots & 18h 50m 36.4s & $-06\arcdeg12\arcmin24\farcs3$ & 17.858 & 0.013 & 17.1555 & 0.0057 & 0.462 &  9.01 & 0.115  &  2\\
4724& \ldots &  18h 50m 35.2s & $-06\arcdeg18\arcmin41\farcs0$ & 18.126 & 0.013 & 17.3207 & 0.0073 & 0.500 & 25.38 & 0.064 &  3\\
4779& \ldots &  18h 50m 43.7s & $-06\arcdeg23\arcmin29\farcs2$ & 18.455 & 0.013 & 17.2955 & 0.0102 & 0.919 &  7.38 & 0.073 &  3\\
4804& \ldots &  18h 50m 42.4s & $-06\arcdeg17\arcmin18\farcs2$ & 17.958 & 0.013 & 17.2570 & 0.0027 & 0.461 & 19.42 & 0.068  &  2\\
5274& \ldots & 18h 51m 24.0s & $-06\arcdeg10\arcmin 48\farcs0$ & 17.972 & 0.012 & 17.3137 & 0.0280 & 0.368 & 42.41 & 0.163 &  3\\
5480& \ldots & 18h 50m 41.5s & $-06\arcdeg12\arcmin07\farcs8$ & 18.290 & 0.016 & 17.4844 & 0.0075 & 0.466 & 25.80 & 0.070 &  3\\
5710& \ldots & 18h 50m 48.7s & $-06\arcdeg11\arcmin34\farcs9$ & 18.249 & 0.014 & 17.6210 & 0.0085 & 0.352 & 7.784 & 0.061 &  3\\
6647& \ldots & 18h 50m 33.5s & $-06\arcdeg21\arcmin48\farcs3$ & 18.395 & 0.018 & 17.6968 & 0.0100 & 0.218 & 40.682 & 0.128 &  3\\
6805& \ldots &  18h 50m 38.1s & $-06\arcdeg19\arcmin53\farcs8$ & 18.799 & 0.021 & 17.8643 & 0.0063 & 0.745 &12.516 & 0.088 &  3\\
6870& \ldots & 18h 51m 18.8s &  $-06\arcdeg13\arcmin 22\farcs4$ & 18.766 & 0.016 & 18.1076 & 0.0117 & 0.419 & 28.16 & 0.113  &  1\\
7467& \ldots &  18h 51m 23.0s & $-06\arcdeg21\arcmin41\farcs2$   & 19.145 & 0.021 & 18.2640 & 0.0114 & 0.640 & 34.06 & 0.107  &  2\\
7522& \ldots &  18h 51m 22.6s & $-06\arcdeg21\arcmin01\farcs9$  & 19.115 & 0.021 & 18.1868 & 0.0043 & 0.688 & 45.02 & 0.141 &  2\\
8066& \ldots &  18h 50m 32.3s & $-06\arcdeg19\arcmin53\farcs4$ & 19.211 & 0.024 & 18.4849 & 0.0269 & 0.636 & 47.64 & 0.263 &  3\\
8146& \ldots & 18h 51m 10.2s &  $-06\arcdeg12\arcmin50\farcs1$ & 19.332 & 0.022 & 18.4508 & 0.0170 & 0.760 & 21.12 & 0.132 &  3\\
8641& \ldots & 18h 51m 09.3s & $-06\arcdeg11\arcmin47\farcs1$ & 19.588 & 0.028 & 18.6337 & 0.0137 & 0.620 & 20.23 & 0.133 &  3\\
\enddata
\label{data}
\tablecomments{The data included in the table columns are as follows:
  (1) variable identification number from this study, (2) variable
  identification number from the \textit{Hubble Space Telescope} Guide
  Star Catalog, variable identification number from \citealt{mps77}
  (membership probability given in parentheses),(3) Right Ascension
  (J2000.0), (4) Declination (J2000.0), (5)mean $V$ apparent
  magnitude, (6) $1\sigma$ error on the mean $V$ apparent magnitude,
  (7) median $R$ apparent magnitude, (8) $1\sigma$ error on the median
  $R$ apparent magnitude, (9) $(V-R)$ color index (shifted to match
  theoretical isochrones; see $\S{\ref{CMD}}$ for details), (10)
  computed value for the Welch-Stetson variability statistic $I_{WS}$,
  (11) computed value for the Stetson-$J$ variability statistic,
  (12)type of variability (see footnote to column 12).}
\tablenotetext{a}{Type of variable star: 1=\ds or
		pulsating variable,2=detached eclipsing binary,3=W UMa
		variable,4=unknown/unidentified variable.}
\end{deluxetable}

\begin{deluxetable}{lccccc}
\tablewidth{0pt}
\tablecaption{Periods, Amplitudes, and Radial Distance from Cluster Center for Detected Variable Stars}
\tablehead{\colhead{ID} & \colhead{GSC,ID$_{\textrm{MPS}}(P\%$)} &  \colhead{Period (d)\tablenotemark{a}} & \colhead{Amplitude (mag)} & \colhead{$r$} & \colhead{T\tablenotemark{b}}}
\startdata
220 & 683(0)             & \ldots     & 0.05  & 3.71 & 4\\
243 & 5126:4687,1268(62) & 0.86577(1) & 0.17  & 1.55 & 3\\
320 & 5126:5601,1711(98) & 0.05453(1) & 0.02  & 5.99 & 1\\
331 & 5126:4823,676(0)   & 0.04521(1) & 0.04  & 3.08 & 1\\
536 & 5126:5659,1531(96) & 0.04201(1) & 0.01  & 2.63 & 1\\
614 & 5126:5664,1237(83) & 0.06522(1) & 0.02  & 2.23 & 1\\
619 &1097(80)            & 0.04154(1) & 0.02  & 5.39 & 1\\
708 & 1521(0) &            \ldots     & 0.15  & 6.63 & 4\\
729 & 900(81) &            \ldots     & 0.20  & 1.32 & 2\\
977 &1596(52) &            \ldots     & 0.10  & 3.71 & 2\\
1026& 1715(7) &            \ldots     & 0.10  & 5.66 & 2\\
1340& \ldots  &            3.79130(1) & 0.55  & 5.02 & 2\\
1583& \ldots  &            1.11763(1) & 0.35  & 8.95 & 2\\
1814& \ldots  &            0.70569(1) & 0.50  & 6.83 & 2\\
1938& \ldots  &            5.62050(1) & 0.50  & 5.59 & 2\\
2119& \ldots  &            \ldots     & 0.10  & 8.53 & 2\\
2406& \ldots  &            \ldots     & 0.40  & 5.95 & 2\\
2740& \ldots  &            0.39464(1) & 0.30  & 2.65 & 3\\
3025& \ldots  &            0.441864(3)& 0.65  & 7.87 & 3\\
3064& \ldots  &            0.44110(1) & 0.20  & 6.37 & 3\\
3096& \ldots  &            1.65174(1) & 0.80  & 4.27 & 2\\
3250& \ldots  &            0.35208(1) & 0.35  & 6.28 & 3\\
3408& \ldots  &            0.25532(1) & 0.20  & 3.18 & 3\\
3597& \ldots  &            0.47349(1) & 0.30  & 5.91 & 3\\
4678& \ldots  &            0.72546(1) & 0.45  & 7.64 & 2\\
4724& \ldots  &            0.41706(1) & 0.20  & 7.32 & 3\\
4779& \ldots  &            0.38382(1) & 0.30  & 8.72 & 3\\
4804& \ldots  &            \ldots     & 0.30  & 5.21 & 2\\
5274& \ldots  &            0.3468800(1)& 0.40  & 7.56 & 3\\
5480& \ldots  &            0.43009(1) & 0.20  & 6.73 & 3\\
5710& \ldots  &            0.46243(1) & 0.20  & 5.84 & 3\\
6647& \ldots  &            0.43220(1) & 0.30  & 9.22 & 3\\
6805& \ldots  &            0.34516(1) & 0.20  & 7.20 & 3\\
6870& \ldots  &            0.08081(1) & 0.40  & 4.89 & 1\\
7467& \ldots  &            \ldots     & 0.30  & 7.43 & 2\\
7522& \ldots  &            \ldots     & 0.80  & 6.89 & 2\\
8066& \ldots  &            0.42260(1) & 0.70  & 8.46 & 3\\
8146& \ldots  &            0.29357(1) & 0.40  & 3.84 & 3\\
8641& \ldots  &            0.45950(1) & 0.30  & 4.72 & 3\\
\enddata
\label{data_pa}
\tablecomments{The data included in the table columns are as follows:
  (1) variable identification number from this study, (2) variable
  identification number from the \textit{Hubble Space Telescope} Guide
  Star Catalog, variable identification number from \citealt{mps77}
  (membership probability given in parentheses), (3) period (in days)
  of detected variability, (4) maximum peak-to-peak amplitude (in
  magnitudes) of detected variability, (5) radial distance (in
  arcminutes) from cluster center, and (6) type of variability (see
  footnote to column 6).}

\tablenotetext{a}{Probable errors (as determined by the precision of
		the period search) are given in parentheses .  In some
		cases insufficent data (lack of multiple events)
		prevented a period determination.}
\tablenotetext{b}{Type of variable star: 1=\ds or
		pulsating variable,2=detached eclipsing binary,3=W UMa
		variable,4=unknown/unidentified variable.}
\end{deluxetable}

\begin{deluxetable}{ll}
\tablewidth{0pt}
\tablecaption{Wilson-Devinney Light Curve Solution for ID 3025}
\tablehead{\colhead{Parameter} & \colhead{Value}}
\startdata
Mass Ratio & 0.417(2)\\
Period & 0.4418638 d\\
Inclination & 82\udeg52(14)\\
Fillout Factor & 0.245\\
Third Light & 0.00 (assumed)\\
Phase Shift & 0.00 (assumed)\\
Wavelength & $6400\mbox{\AA}$\\
\\
$\Omega_{1}$ & 2.6477(48)\\
r$_{1}$(pole) & 0.44127(118)\\
r$_{1}$(side) & 0.47349(163)\\
r$_{1}$(back) & 0.50528(233)\\
T$_{1}$ & 7300 K (assumed)\\
L$_{1}$ & 0.6956(12)\\
g$_{1}$ & 1.00 (assumed)\\
x$_{1}$ & 0.405 (assumed)\\
A$_{1}$ & 1.00 (assumed)\\
\\
$\Omega_{2}$ & 2.6477(48)\\
r$_{2}$(pole) & 0.29830(99)\\
r$_{2}$(side) & 0.31296(125)\\
r$_{2}$(back) & 0.35532(252)\\
T$_{2}$ & 7196(7) K \\
L$_{2}$ & 0.3044\\
g$_{2}$ & 1.00 (assumed)\\
x$_{2}$ & 0.405 (assumed)\\
A$_{2}$ & 1.00 (assumed)\\
\enddata
\label{3025_table}
\tablecomments{Parameter descriptions: Fillout Factor=parametric
characterization of the equipotential surface (percentage that the
surface potential lies between the inner and outer Lagrangian
surfaces); $\Omega_{1,2}$=parametric characterization of the
gravitational equipotential surface; $r_{1,2}$(pole,side,back)=radii
along differing axes of the system; $T_{1,2}$=effective surface
temperature;$L_{1,2}$=fractional luminosity; $g_{1,2}$=gravity
darkening (brightening) exponent; $x_{1,2}$=limb darkening coefficient
(from \citealt{va93}); $A_{1,2}$=bolometric albedo (reflection
coefficent).}
\end{deluxetable}

\begin{deluxetable}{lll}
\tablewidth{0pt} \tablecaption{Light Curve Solutions for ID 5274}
\tablehead{\colhead{Parameter} & \colhead{\textit{Binary Maker 3}} & \colhead{Wilson-Devinney}} 
\startdata 
Mass Ratio & 0.25 & 0.236(6)\\ 
Period & 0.3468800 d & 0.3468800 d\\ 
Inclination & 78\udeg0 & 76\udeg16(75)\\
Fillout Factor & 0.25 & 0.468\\ 
Third Light & 0.00 & 0.00 (assumed)\\ 
Phase Shift & 0.50 & 0.50 (assumed)\\ 
Wavelength &  $6400\mbox{\AA}$ & $6400\mbox{\AA}$\\
\\
$\Omega_{1}$ & 2.3135 & 2.2501(75)\\
r$_{1}$(pole) & 0.47892  & 0.49062(293)\\
r$_{1}$(side) & 0.52012 & 0.53639(441)\\
r$_{1}$(back) & 0.54701 & 0.56554(608)\\
T$_{1}$ & 6650 K & 6550 K (assumed)\\
L$_{1}$ & 0.7377 & 0.7462(64)\\
g$_{1}$ & 0.32 & 0.32 (assumed)\\
x$_{1}$ & 0.38 & 0.38 (assumed)\\
A$_{1}$ & 0.50 & 0.50 (assumed)\\
\\
$\Omega_{2}$ & 2.3135 & 2.2501(75)\\
r$_{2}$(pole)& 0.25770 & 0.26200(503)\\
r$_{2}$(side)& 0.26953 & 0.27534(635)\\
r$_{2}$(back)& 0.31077 & 0.32720(1586)\\
T$_{2}$ & 7000 K & 6995(22) K\\
L$_{2}$ & 0.2623 & 0.2538\\
g$_{2}$ & 0.32 & 0.32 (assumed)\\
x$_{2}$ & 0.38 & 0.38 (assumed)\\
A$_{2}$ & 0.50 & 0.50 (assumed)\\
\enddata
\tablecomments{Parameter descriptions: Fillout Factor=parametric
characterization of the equipotential surface (percentage that the
surface potential lies between the inner and outer Lagrangian
surfaces); $\Omega_{1,2}$=parametric characterization of the
gravitational equipotential surface; $r_{1,2}$(pole,side,back)=radii
along differing axes of the system; $T_{1,2}$=effective surface
temperature;$L_{1,2}$=fractional luminosity; $g_{1,2}$=gravity
darkening (brightening) exponent; $x_{1,2}$=limb darkening coefficient
(from \citealt{va93}); $A_{1,2}$=bolometric albedo (reflection
coefficent).}
\label{5274_table}
\end{deluxetable}

\end{document}